\documentclass{JHEP3}

\usepackage{amsmath}
\usepackage{amsfonts}
\usepackage{amssymb}
\usepackage{amscd}
\usepackage{graphicx}
\usepackage{citesort}
\usepackage{mathrsfs}
\usepackage{bbm}

\def\S{{\mathbb S}}
\def\V{{\mathscr V}}

\def\a{\alpha}

\newcommand{\alg}[1]{\mathfrak{#1}}
\newcommand{\su}{\alg{su}}

\newcommand{\psu}{\alg{psu}}

\newcommand{\stateA}[1]{|#1\rangle^{\rm{I}}}
\newcommand{\stateB}[1]{|#1\rangle^{\rm{II}}}
\newcommand{\stateC}[1]{|#1\rangle^{\rm{III}}}
\newcommand{\costateA}[1]{ ^{\rm{I}}\langle #1}

\def\ads{{\rm AdS}_5\times {\rm S}^5}

\author{G Arutyunov\footnote{ Correspondent fellow at
Steklov Mathematical Institute, Moscow.}, M de Leeuw, A Torrielli
\footnote{E-mail: G.E.Arutyunov,M.deLeeuw,A.Torrielli@uu.nl}
 \\  {\it Institute for Theoretical
Physics and Spinoza Institute,\\ Utrecht University, 3508 TD
Utrecht, The Netherlands}}

\abstract{We determine the S-matrix that describes scattering of
arbitrary bound states in the light-cone string theory in $\ads$.
The corresponding construction relies on the Yangian symmetry and
the superspace formalism for the bound state representations. The
basic analytic structure supporting the S-matrix entries turns out
to be the hypergeometric function $_4\hspace{-0.03cm}F_3$. We show
that for particular bound state numbers it reproduces all the
scattering matrices previously obtained in the literature. Our
findings should be relevant for the TBA and L\"uscher approaches
to the finite-size spectral problem. They also shed some light on
the construction of the universal R-matrix for the
centrally-extended $\psu(2|2)$ superalgebra.

}

\title{The Bound State S-Matrix for $\ads$ Superstring}

\preprint{
          \tiny{ITP-UU-09-06}\\[-.5ex]
          \tiny{SPIN-09-06}\\[-.5ex]
          }
\begin{document}

\section{Introduction}
The S-matrix approach has been recently recognized as an
indispensable tool to study the spectral problem of both the
$\ads$ superstring and the dual gauge theory
\cite{Minahan:2002ve,Beisert:2004hm,Arutyunov:2004vx,Staudacher:2004tk,Beisert:2005fw,Beisert:2005tm,Beisert:2006ib,Beisert:2006ez,Klose:2006zd,Arutyunov:2009ga}.
In the uniform light-cone gauge
\cite{Arutyunov:2004yx,Frolov:2006cc} the corresponding string
sigma model describes a two-dimensional massive quantum field
theory of eight bosons and eight fermions. The light-cone momentum
$P_+$, which is a gauge fixing parameter, plays the role of the
string length. In the limit when $P_+$ tends to infinity, the
sigma model is defined on a plane, which calls for the application
of scattering theory.


For a scattering process, the statement of integrability implies
absence of particle production and factorization of multi-particle
scattering into a sequence of two-body events. Since the sigma
model does not have two-dimensional Lorentz invariance, the
two-body S-matrix for scattering of fundamental particles (giant
magnons \cite{Hofman:2006xt}) has a rather intricate structure. It
is invariant w.r.t. the superalgebra $\psu(2|2)\oplus \psu(2|2)$
enhanced by three central charges depending on particle momenta
\cite{Beisert:2005tm}. The latter algebra is a symmetry of the
light-cone gauge fixed Hamiltonian in the off-shell theory where
the level-matching condition, i.e. the requirement of vanishing of
the total world-sheet momentum, is suspended
\cite{Arutyunov:2006ak}. The matrix structure of the two-body
S-matrix appears to be uniquely fixed by the centrally-extended
$\psu(2|2)\oplus \psu(2|2)$ algebra, the Yang-Baxter equation and
the generalized physical unitarity condition
\cite{Beisert:2005tm,Arutyunov:2006yd,Arutyunov:2007tc}, while the
overall scaling factor is severely constrained by crossing
symmetry \cite{Janik:2006dc}.


In the limit of infinite light-cone momentum, in addition to the
fundamental particles, the spectrum of the string sigma model
contains an infinite tower of bound states \cite{Dorey:2006dq}.
 More explicitly, $\ell$-particle
bound states comprise into the tensor product of two $4\ell$-dim
atypical totally symmetric multiplets of the centrally-extended
$\psu(2|2)$ algebra  \cite{Beisert:2006qh}.


Concerning the finite $P_+$ spectrum, any power-like $1/P_+$
corrections to the energy of a multi-particle state can be
obtained by means of the Bethe-Yang equations
\cite{Beisert:2005fw,Martins:2007hb,Leeuw:2007uf} representing the
quantization condition for the particle momenta.  A complete
handle on the string asymptotic spectrum and the associated
Bethe-Yang equations requires, in principle, the knowledge of the
$\psu(2|2)$-invariant S-matrices $\mathbb{S}^{\ell_1\ell_2}$ which
describe the scattering of $\ell_1$- and $\ell_2$-particle bound
states.


In addition to the power-like corrections to the spectrum, there
are also exponentially small corrections which are not captured by
the Bethe-Yang equations. The leading exponential corrections
\cite{Arutyunov:2006gs,Klose:2008rx,Minahan:2008re} to the
dispersion relation for the fundamental particles  and the bound
states can be derived by applying the perturbative L\"uscher's
approach \cite{Luscher:1985dn}, which is also based on the
knowledge of the world-sheet S-matrix
\cite{Ambjorn:2005wa,Janik:2007wt,Hatsuda:2008gd,Gromov:2008ie,Heller:2008at}.
It appears that  L\"uscher's approach could also be applied to
find perturbative scaling dimensions of gauge theory operators up
to the first order where the Bethe-Yang description breaks down
\cite{Bajnok:2008bm,Fiamberti:2007rj} (see also
\cite{Beccaria:2009eq}). The energy of the two-particle state
corresponding to the Konishi operator receives contribution from
the whole tower of $\ell$-particle bound states which travel
around a cylinder of finite circumference $P_+$. The corresponding
computation thus exploits the scattering matrices of fundamental
particles with bound states, $\mathbb{S}^{1 \ell}$. A more general
treatment of the scaling dimension of a gauge theory operator
corresponding to an on-shell bound state particle will thus
require the knowledge of a generic scattering matrix
$\mathbb{S}^{\ell_1\ell_2}$.


To find the exact, i.e. non-perturbative string theory spectrum,
one could try to adapt the Thermodynamic Bethe Ansatz (TBA)
approach, originally developed for relativistic integrable models
\cite{Zamolodchikov:1989cf}. To this end, one has to find the
scattering matrices for bound states of the accompanying mirror
theory \cite{Arutyunov:2007tc}. The bound states of this theory
are ``mirror reflections" of those of the original string model.
Furthermore, the scattering matrix of the mirror particles is
obtained from $\S^{\ell_1\ell_2}$ by double Wick rotation and the
transfer to the anti-symmetric representations. Although the
Bethe-Yang equations for bound states of the original and the
mirror model can be obtained by fusing the Bethe equations for the
corresponding fundamental particles\footnote{We refer to
\cite{deLeeuw:2008ye} for a derivation of the corresponding Bethe
equations from Yangian symmetry. See the references in
\cite{Torrielli:2008wi} for an earlier literature on Yangians in
AdS/CFT. Recent progress can be found in
\cite{Zwiebel:2008gr,Bargheer:2008jt}.}, the knowledge of
$\mathbb{S}^{\ell_1\ell_2}$ might be relevant for an alternative
approach based on functional relations between eigenvalues of the
transfer matrices. In particular, it would be interesting to check
various conjectures about such eigenvalues in higher rank
representations. For an interesting recent development in this
direction we refer to
\cite{Kazakov:2007fy,Gromov:2008gj,Arutyunov:2009zu,Gromov:2009tv}.


The aim of this paper is to complete the program of
\cite{Arutyunov:2008zt} by finding the S-matrix
$\mathbb{S}^{\ell_1\ell_2}$ which corresponds to scattering of
bound states with arbitrary bound state numbers. This will be done
by using the Yangian symmetry \cite{Beisert:2007ds} in conjunction
with the superspace formalism \cite{Arutyunov:2008zt}.

According to \cite{Arutyunov:2008zt},  the $\ell$-particle bound
state representation $\V_\ell$ of the centrally extended
$\su(2|2)$ algebra can be realized on the space of homogeneous
(super)symmetric polynomials of degree $\ell$ depending on two
bosonic and two fermionic variables, $w_a$ and $\theta_{\a}$,
respectively. Thus, the representation space is spanned by an
irreducible short superfield $\Phi_\ell(w,\theta)$. In this
realization the algebra generators are represented by differential
operators linear in ${\mathbb J}$ in $w_a$ and $\theta_{\a}$
 with coefficients depending on the representation
parameters (the particle momenta). The S-matrix
$\mathbb{S}^{\ell_1\ell_2}$ acts on the product of two superfields
$\Phi_{\ell_1}(w,\theta)\Phi_{\ell_2}(v,\vartheta)$ as a
differential operator of degree $\ell_1+\ell_2$. We require this
operator to obey the following intertwining property
\begin{eqnarray}
\nonumber \S^{\ell_1\ell_2}\cdot {\mathbb J}_{12}= {\mathbb
J}_{21}\cdot \S^{\ell_1\ell_2}\, ,
\end{eqnarray}
where ${\mathbb J}_{12}$ and ${\mathbb J}_{21}$ are the generators
of the centrally extended $\psu(2|2)$ in the corresponding
two-particle representation. For $\su(2)$ subalgebras of
$\psu(2|2)$ this condition literally means the invariance of the
S-matrix, while for the supersymmetry generators it involves the
braiding (non-local) factors
\cite{Gomez:2006va,Plefka:2006ze,Arutyunov:2006yd} to be discussed
later. In \cite{Arutyunov:2008zt} the lower-dimensional examples
$\S^{11}$, $\S^{12}$ and $\S^{22}$ have been found by solving the
above invariance condition together with the Yang-Baxter equation.
The fundamental S-matrix $\S^{11}$ has been shown
\cite{Beisert:2007ds} to commute with the Yangian for the
centrally extended $\psu(2|2)$ algebra. It appears that the new
examples $\S^{12}$ and $\S^{22}$ also respect Yangian symmetry
which provides an alternative to solving the Yang-Baxter equation
\cite{deLeeuw:2008dp}. Denoting by $\hat{\mathbb{J}}$ a Yangian
generator acting in the finite-dimensional evaluation
representation corresponding to a bound state, we thus require
\begin{eqnarray}
\nonumber \S^{\ell_1\ell_2}\cdot \hat{{\mathbb J}}_{12}=
\hat{{\mathbb J}}_{21}\cdot \S^{\ell_1\ell_2}\, ,
\end{eqnarray}
for any $\S^{\ell_1\ell_2}$. The Yangian carries a natural Hopf
algebra structure \cite{Beisert:2007ds}, so that $\hat{{\mathbb
J}}_{12}$ and $\hat{{\mathbb J}}_{21}$ can be regarded as the
coproducts $\Delta(\hat{{\mathbb J}})$ and
$\Delta^{op}(\hat{{\mathbb J}})$, respectively, evaluated on bound
state representations. Here $\Delta^{op}$ stands for the opposite
coproduct.


Our construction of $\S^{\ell_1\ell_2}$ will involve three
two-particle bases: the first is the standard one corresponding to
the product $\Phi_{\ell_1}(w,\theta)\Phi_{\ell_2}(v,\vartheta)$,
the second is given by certain products
$\Delta(\mathbb{J}^{a_1})\cdots \Delta(\hat{\mathbb{J}}^{a_k})$
and the third by $\Delta^{op}(\mathbb{J}^{a_1})\cdots
\Delta^{op}(\hat{\mathbb{J}}^{a_k})$. Denoting by $\Lambda$ and
$\Lambda^{op}$ the transition matrices from the second and the
third basis to the standard one, we will show that the matrix
representation of the $\S$-operator in the standard basis is
nothing else but
\begin{eqnarray}
\nonumber \S^{\ell_1\ell_2}=\Lambda^{op}\cdot \Lambda^{-1}\, .
\end{eqnarray}
Finding $\Lambda$ and $\Lambda^{op}$, we will thus establish the
matrix form of $\S^{\ell_1\ell_2}$. As we will see, the basic
analytic structure supporting the expressions for the
corresponding matrix elements is given by the hypergeometric
function $_4\hspace{-0.03cm}F_3$. We point out that our
construction of $\S$ involves linear algebra only and it can be
easily implemented in the Mathematica program. It is interesting
to point out that the above matrix factorization of the S-matrix
reminds of the Borel decomposition of the double of the Yangian.


Taking the tensor product of two $\psu(2|2)$-invariant S-matrices,
one obtains $\psu(2|2)\oplus\psu(2|2)$-invariant world-sheet
S-matrices which describe the scattering of bound states in the
light-cone string theory on $\ads$. Of course, the resulting
S-matrix should be multiplied with the proper overall scalar
factor obeying crossing symmetry. This factor has been already
obtained in \cite{Chen:2006gq,Roiban:2006gs,Arutyunov:2008zt}.


Having found the general bound state scattering matrix
$\S^{\ell_1\ell_2}$, we then verify that it reproduces all the
previously obtained special cases and satisfies the necessary
physicality requirements. In particular, we investigated its
analytic structure and confirmed that it only exhibits the
expected physical pole corresponding to the formation of a bound
state of rank $\ell_1+\ell_2$ . This completes the bound state
S-matrix program.


It is worth stressing that, besides the uses in the TBA or
L\"uscher's approaches , which are physically the most relevant
ones, there is also a mathematical interest in deriving the
universal R/S-matrix for the centrally-extended $\psu(2|2)$
algebra. The solution to this problem has been so far elusive, in
spite of the progress nevertheless achieved
\cite{Moriyama:2007jt,Matsumoto:2007rh,Beisert:2007ty,Heckenberger:2007ry,Spill:2008tp,Matsumoto:2008ww}.
The present work provides an important step towards finding the
universal R-matrix. In fact, our formulae present the explicit
R-matrix entries for arbitrary evaluation representations of the
relevant Yangian, namely, all (finite-dimensional) bound-state
representations. On one hand, this exhausts a large class of
representations. On the other hand, it may now become easier to
establish a universal form of the R-matrix which could reproduce
all these expressions, as started in \cite{Torrielli:2008wi}.


The paper is organized as follows. In the next section we
introduce our notations and formulate our main result -- the
S-matrix $\S^{\ell_1\ell_2}$ of arbitrary bound states in the
standard basis. The rest of the paper contains details of the
derivation and verifications of various properties. Discussion of
certain analytic aspects is relegated to the appendices.


\section{Kinematical Structure of the S-Matrix}
In this section, we will discuss the kinematical structure of the
S-matrix. In particular, we will use $\alg{su}(2|2)$ invariance to
show that the S-matrix is of block diagonal form.

\subsection{Centrally extended $\alg{su}(2|2)$}
We will first discuss centrally extended $\mathfrak{su}(2|2)$.
This algebra has bosonic generators $\mathbb{R},\mathbb{L}$,
supersymmetry generators $\mathbb{Q},\mathbb{G}$ and central
charges $\mathbb{H},\mathbb{C},\mathbb{C}^{\dag}$. The non-trivial
commutation relations between the generators are given by
\begin{eqnarray}
\begin{array}{lll}
\ [\mathbb{L}_{a}^{\ b},\mathbb{J}_{c}] = \delta_{c}^{b}\mathbb{J}_{a}-\frac{1}{2}\delta_{a}^{b}\mathbb{J}_{c}, &\qquad & \ [\mathbb{R}_{\alpha}^{\ \beta},\mathbb{J}_{\gamma}] = \delta_{\gamma}^{\beta}\mathbb{J}_{\alpha}-\frac{1}{2}\delta_{\alpha}^{\beta}\mathbb{J}_{\gamma},\\
\ [\mathbb{L}_{a}^{\ b},\mathbb{J}^{c}] = -\delta_{a}^{c}\mathbb{J}^{b}+\frac{1}{2}\delta_{a}^{b}\mathbb{J}^{c}, &\qquad& \ [\mathbb{R}_{\alpha}^{\ \beta},\mathbb{J}^{\gamma}] = -\delta^{\gamma}_{\alpha}\mathbb{J}^{\beta}+\frac{1}{2}\delta_{\alpha}^{\beta}\mathbb{J}^{\gamma},\\
\ \{\mathbb{Q}_{\alpha}^{\ a},\mathbb{Q}_{\beta}^{\
b}\}=\epsilon_{\alpha\beta}\epsilon^{ab}\mathbb{C},&\qquad&\ \{\mathbb{G}^{\ \alpha}_{a},\mathbb{G}^{\ \beta}_{b}\}=\epsilon^{\alpha\beta}\epsilon_{ab}\mathbb{C}^{\dag},\\
\ \{\mathbb{Q}_{\alpha}^{a},\mathbb{G}^{\beta}_{b}\} =
\delta_{b}^{a}\mathbb{R}_{\alpha}^{\ \beta} +
\delta_{\alpha}^{\beta}\mathbb{L}_{b}^{\ a}
+\frac{1}{2}\delta_{b}^{a}\delta_{\alpha}^{\beta}\mathbb{H}.&&
\end{array}
\end{eqnarray}
The eigenvalues of the central charges are denoted by
$H,C,C^{\dag}$. The charge $\mathbb{H}$ is Hermitian and the
charges $\mathbb{C},\mathbb{C}^{\dag}$ and the generators
$\mathbb{Q},\mathbb{G}$ are conjugate to each other.\smallskip

For computational purposes, it proves worthwhile to consider
representations of the algebra in the superspace formalism.
Consider the vector space of analytic functions of two bosonic
variables $w_{1,2}$ and two fermionic variables $\theta_{3,4}$.
Since we are dealing with analytic functions we can expand any
such function $\Phi(w,\theta)$:
\begin{eqnarray}
\Phi(w,\theta) &=&\sum_{\ell=0}^{\infty}\Phi_{\ell}(w,\theta),\nonumber\\
\Phi_{\ell} &=& \phi^{a_{1}\ldots a_{\ell}}w_{a_{1}}\ldots
w_{a_{\ell}} +\phi^{a_{1}\ldots a_{\ell-1}\alpha}w_{a_{1}}\ldots
w_{a_{\ell-1}}\theta_{\alpha}+\nonumber\\
&&\phi^{a_{1}\ldots a_{\ell-2}\alpha\beta}w_{a_{1}}\ldots
w_{a_{\ell-2}}\theta_{\alpha}\theta_{\beta}.
\end{eqnarray}
The representation that describes $\ell$-particle bound states is
$4\ell$ dimensional. It is realized on a graded vector space with
basis $|e_{a_{1}\ldots a_{\ell}}\rangle, |e_{a_{1}\ldots
a_{\ell-1}\alpha}\rangle,|e_{a_{1}\ldots
a_{\ell-2}\alpha\beta}\rangle$, where $a_{i}$ are bosonic indices
and $\alpha,\beta$ are fermionic indices, and each of the basis
vectors is totally symmetric in the bosonic indices and
anti-symmetric in the fermionic indices
\cite{Arutyunov:2008zt,Beisert:2006qh,Dorey:2006dq}. In terms of
the above analytic functions, the basis vectors of the totally
symmetric representation can evidently be identified as
$|e_{a_{1}\ldots a_{\ell}}\rangle \leftrightarrow w_{a_{1}}\ldots
w_{a_{\ell}},|e_{a_{1}\ldots a_{\ell-1}\alpha}\rangle
\leftrightarrow w_{a_{1}}\ldots w_{a_{\ell-1}}\theta_{\alpha}$ and
$|e_{a_{1}\ldots a_{\ell-1}\alpha\beta}\rangle \leftrightarrow
w_{a_{1}}\ldots w_{a_{\ell-2}}\theta_{\alpha}\theta_{\beta}$,
respectively. In other words, we find the atypical totally
symmetric representation describing $\ell$-particle bound states
when we restrict to terms $\Phi_{\ell}$.
\smallskip

In this representation the algebra generators can be written in
differential operator form as
\begin{eqnarray}\label{eqn;AlgDiff}
\begin{array}{lll}
  \mathbb{L}_{a}^{\ b} = w_{a}\frac{\partial}{\partial w_{b}}-\frac{1}{2}\delta_{a}^{b}w_{c}\frac{\partial}{\partial w_{c}}, &\qquad& \mathbb{R}_{\alpha}^{\ \beta} = \theta_{\alpha}\frac{\partial}{\partial \theta_{\beta}}-\frac{1}{2}\delta_{\alpha}^{\beta}\theta_{\gamma}\frac{\partial}{\partial \theta_{\gamma}}, \\
  \mathbb{Q}_{\alpha}^{\ a} = a \theta_{\alpha}\frac{\partial}{\partial w_{a}}+b\epsilon^{ab}\epsilon_{\alpha\beta} w_{b}\frac{\partial}{\partial \theta_{\beta}}, &\qquad& \mathbb{G}_{a}^{\ \alpha} = d w_{a}\frac{\partial}{\partial \theta_{\alpha}}+c\epsilon_{ab}\epsilon^{\alpha\beta} \theta_{\beta}\frac{\partial}{\partial w_{b}},
\end{array}
\end{eqnarray}
and the central charges are
\begin{eqnarray}
\begin{array}{ll}
 \mathbb{C} = ab \left(w_{a}\frac{\partial}{\partial w_{a}}+\theta_{\alpha}\frac{\partial}{\partial
 \theta_{\alpha}}\right),& \mathbb{C}^{\dag} = cd \left(w_{a}\frac{\partial}{\partial w_{a}}+\theta_{\alpha}\frac{\partial}{\partial
 \theta_{\alpha}}\right),\\
 \mathbb{H}= (ad +bc)\left(w_{a}\frac{\partial}{\partial w_{a}}+\theta_{\alpha}\frac{\partial}{\partial
 \theta_{\alpha}}\right).
\end{array}
\end{eqnarray}
To form a representation, the parameters $a,b,c,d$ must satisfy
the condition $ad-bc=1$. The central charges become $\ell$
dependent:
\begin{eqnarray}
H= \ell(ad+bc),\qquad C =\ell ab , \qquad C^{\dag} =\ell cd.
\end{eqnarray}
The parameters $a,b,c,d$ can be expressed in terms of the particle
momentum $p$ and the coupling $g$:
\begin{eqnarray}
\begin{array}{lll}
  a = \sqrt{\frac{g}{2\ell}}\eta, & \quad & b = \sqrt{\frac{g}{2\ell}}
\frac{i\zeta}{\eta}\left(\frac{x^{+}}{x^{-}}-1\right), \\
  c = -\sqrt{\frac{g}{2\ell}}\frac{\eta}{\zeta x^{+}}, & \quad &
d=\sqrt{\frac{g}{2\ell}}\frac{x^{+}}{i\eta}\left(1-\frac{x^{-}}{x^{+}}\right),
\end{array}
\end{eqnarray}
where the parameters $x^{\pm}$ satisfy
\begin{eqnarray}
x^{+} +
\frac{1}{x^{+}}-x^{-}-\frac{1}{x^{-}}=\frac{2i\ell}{g},\qquad
\frac{x^{+}}{x^{-}} = e^{ip}
\end{eqnarray}
and the parameters $\eta$ are given by
\begin{eqnarray}\label{eqn;ScatteringBasis}
\eta = e^{i\xi}\eta(p),\qquad \eta(p)=
e^{i\frac{p}{4}}\sqrt{ix^{-}-ix^{+}},\qquad \zeta = e^{2i\xi}.
\end{eqnarray}
The fundamental representation, which is used in the derivation of
the S-matrix scattering two fundamental multiplets
\cite{Beisert:2005tm,Arutyunov:2006yd}, is obtained by taking
$\ell=1$.

The bound state S-matrices should respect the $\alg{su}(2|2)$
symmetry, by requiring invariance under the coproducts of the
generators
\begin{eqnarray}\label{eqn;SymmPropSU22}
\S~\Delta(\mathbb{J}^{A})&=&\Delta^{op}(\mathbb{J}^{A})~\S,
\end{eqnarray}
where $\Delta^{op} = P \Delta$, with $P$ the graded
permutation\footnote{ We remind that, in the non-local formalism
of \cite{Arutyunov:2006yd}, one needs to explicitly permute the
two representations when acting with $P$.}.

\subsection{Invariant subspaces}

Consider two bound states with bound state numbers $\ell_1,\ell_2$
respectively. The tensor product of the corresponding bound state
representations in superspace is given by:
\begin{eqnarray}
\Phi(w,\theta)\Phi(v,\vartheta),
\end{eqnarray}
where $w,\theta$ denote the superspace variables of the first
particle and $v,\vartheta$ describe the representation of the
second particle.

The S-matrix acts on this tensor space and should, according to
(\ref{eqn;SymmPropSU22}), commute with $\Delta(\mathbb{L})^1_1$
and $\Delta(\mathbb{R})^3_3$. From this, it is easily deduced that
the numbers
\begin{eqnarray}
K^{\rm{II}} &=&
\#\theta_3+\#\theta_4+\#\vartheta_3+\#\vartheta_4+2\#w_2+2\#v_2,\nonumber\\
K^{\rm{III}} &=& \#\theta_3+\#\vartheta_3+\#w_2+\#v_2
\end{eqnarray}
are conserved. The variables $w_2,v_2$ can be interpreted as being
a combined state of two fermions of different type
\cite{Beisert:2005tm}. Hence, the number $K^{\rm{II}}$ corresponds
to the total number of fermions, and the number $K^{\rm{III}}$
counts the number of fermions of one species, say, of type 3. The
fact that these numbers are conserved allows us to define
invariant subspaces, for each of which we will derive the
corresponding S-matrix.\smallskip

Let us write out the tensor product more explicitly. Since we are
considering bound states with bound state number $\ell_1,\ell_2$
we restrict to
\begin{eqnarray}
&&( w_1^{\ell_1-k} w_2^k +  \theta_{3}w_1^{\ell_1-k-1} w_2^k +
\theta_{4}w_1^{\ell_1-k-1} w_2^k +
 \theta_{3}\theta_{4}w_1^{\ell_1-k-1} w_2^{k-1})\times\nonumber\\
&&\times~( v_1^{\ell_2-l} v_2^l + \vartheta_{3}v_1^{\ell_2-l-1}
v_2^l +  \vartheta_{4}v_1^{\ell_2-l-1} v_2^l +
\vartheta_{3}\vartheta_{4}v_1^{\ell_2-l-1} v_2^{l-1}).
\end{eqnarray}
The ranges over which the labels $k,l$ are allowed to vary can be straightforwardly read
off for each term. By multiplying everything out, we reproduce all the basis vectors.
One can compute the quantum numbers $K^{\rm{II}},K^{\rm{III}}$ for
any of these basis vectors. The results are listed in Table
\ref{Tab;Tensor}.
\begin{table}[htbp]
    \centering
        \begin{tabular}{|l|l||l|l||l||l|}
            \hline
            Space 1 & Space 2 &  $K^{\rm{II}}$ & $K^{\rm{III}}$ & $N$ & Case\\
            \hline \hline
            $\theta_{3}w_1^{\ell_1-k-1} w_2^k$ & $\vartheta_{3}v_1^{\ell_2-l-1} v_2^l$ & $2(k+l)+2$ & $k+l+2$ & $k+l$ & Ia\\
            & & & & &\\
            $\theta_{4}w_1^{\ell_1-k-1} w_2^k$ & $\vartheta_{4}v_1^{\ell_2-l-1} v_2^l$ & $2(k+l)+2$ & $k+l$ & $k+l$ & Ib\\
            & & & & &\\
            $\theta_{3}w_1^{\ell_1-k-1} w_2^k$ & $v_2^{\ell_2-l} v_2^l$ & $2(k+l)+1$ & $k+l+1$ & $k+l$ & IIa\\
            $w_1^{\ell_1-k} w_2^k$ & $\vartheta_{3}v_1^{\ell_2-l-1} v_2^l$ & $2(k+l)+1$ & $k+l+1$ & $k+l$ & IIa\\
            $\theta_{3}w_1^{\ell_1-k-1} w_2^k$ & $\vartheta_{3}\vartheta_{4} v_1^{\ell_2-l-1} v_2^{l-1}$ & $2(k+l)+1$ & $k+l+1$ & $k+l$ & IIa\\
            $\theta_{3}\theta_{4}w_1^{\ell_1-k-1} w_2^{k-1}$ & $\vartheta_{3}v_1^{\ell_2-l-1} v_2^l$ & $2(k+l)+1$ & $k+l+1$ & $k+l$ & IIa\\
            & & & & &\\
            $\theta_{4}w_1^{\ell_1-k-1} w_2^k$ & $v_2^{\ell_2-l} v_2^l$ & $2(k+l)+1$ & $k+l$ & $k+l$ & IIb\\
            $w_1^{\ell_1-k} w_2^k$ & $\vartheta_{4}v_1^{\ell_2-l-1} v_2^l$ & $2(k+l)+1$ & $k+l$ & $k+l$ & IIb\\
            $\theta_{4}w_1^{\ell_1-k-1} w_2^k$ & $\vartheta_{3}\vartheta_{4} v_1^{\ell_2-l-1} v_2^{l-1}$ & $2(k+l)+1$ & $k+l$ & $k+l$ & IIb\\
            $\theta_{3}\theta_{4}w_1^{\ell_1-k-1} w_2^{k-1}$ & $\vartheta_{4}v_1^{\ell_2-l-1} v_2^l$ & $2(k+l)+1$ & $k+l$ & $k+l$ & IIb\\
            & & & & &\\
            $w_1^{\ell_1-k} w_2^k$ & $v_2^{\ell_2-l} v_2^l$ & $2(k+l)$ & $k+l$ & $k+l$ & III\\
            $w_1^{\ell_1-k} w_2^k$ & $\vartheta_{3}\vartheta_{4} v_1^{\ell_2-l-1} v_2^{l-1}$ & $2(k+l)$ & $k+l$ & $k+l$ & III\\
            $\theta_{3}\theta_{4}w_1^{\ell_1-k-1} w_2^{k-1}$ & $v_2^{\ell_2-l} v_2^l$ & $2(k+l)$ & $k+l$ & $k+l$ & III\\
            $\theta_{3}\theta_{4}w_1^{\ell_1-k-1} w_2^{k-1}$ & $\vartheta_{3}\vartheta_{4} v_1^{\ell_2-l-1} v_2^{l-1}$ & $2(k+l)$ & $k+l$ & $k+l$ & III\\
            $\theta_{3}w_1^{\ell_1-k-1} w_2^k$ & $\vartheta_{4}v_1^{\ell_2-l-1} v_2^l$ & $2(k+l+1)$ & $k+l+1$ & $k+l+1$ & III\\
            $\theta_{4}w_1^{\ell_1-k-1} w_2^k$ & $\vartheta_{3}v_1^{\ell_2-l-1} v_2^l$ & $2(k+l+1)$ & $k+l+1$ & $k+l+1$ & III\\
            \hline
        \end{tabular}
    \caption{The $16\ell_1\ell_2$ vectors from the tensor product and their $\alg{su}(2)\times\alg{su}(2)$ quantum numbers.}
    \label{Tab;Tensor}
\end{table}
When we take a closer look at the result, we see that when we
order the states by the quantum numbers $K^{\rm{II}},K^{\rm{III}}$,
there are exactly five different types of states:
\begin{description}
  \item[Case Ia:] $K^{\mathrm{II}}=2 N+2, K^{\mathrm{III}}=N+2$,
  \item[Case Ib:] $K^{\mathrm{II}}=2 N+2, K^{\mathrm{III}}=N$,
  \item[Case IIa:] $K^{\mathrm{II}}=2 N+1, K^{\mathrm{III}}=N+1$,
  \item[Case IIb:] $K^{\mathrm{II}}=2 N+1, K^{\mathrm{III}}=N$,
  \item[Case III:] $K^{\mathrm{II}}=2 N, K^{\mathrm{III}}=N$,
\end{description}
for some integer $N$. Each of these states have different quantum
numbers $K^{\rm{II}},K^{\rm{III}}$, hence the states belonging to each
of these cases form a subspace which is left invariant by the
S-matrix.\smallskip

Clearly, vectors from Case Ia and Case Ib only differ by the
exchange $3\leftrightarrow 4$, which is easily realized in terms
of the (fermionic) $\mathfrak{su}(2)$ symmetry generators of type
$\mathbb{R}$. Hence, the subspaces spanned by the two types of
states are isomorphic, and scatter via the same S-matrix. An
analogous relationship connects Case IIa and IIb. Thus, there are
only three non-equivalent cases:
\begin{description}
  \item[Case I:] $K^{\mathrm{II}}=2 N+2, K^{\mathrm{III}}=N+2$,
  \item[Case II:] $K^{\mathrm{II}}=2 N+1, K^{\mathrm{III}}=N+1$,
  \item[Case III:] $K^{\mathrm{II}}=2 N, K^{\mathrm{III}}=N$.
\end{description}
For fixed $N$ (i.e. for fixed $K^{\rm{II}},K^{\rm{III}}$) we
denote the vector spaces, spanned by vectors from each of the
inequivalent cases, by $V^{\rm{I}},V^{\rm{II}},V^{\rm{III}}$
respectively.

\subsection{Basis and relations}

Later on we will introduce different bases for the different
cases, but in this section we will discuss the basis as obtained
by multiplying out the tensor product as seen from Table
\ref{Tab;Tensor}. We will call this basis the standard basis.

\subsection*{Case I, $K^{\mathrm{II}}=2 N +2, K^{\mathrm{III}}=N+2$.}

For fixed $N$, the vector space of states $V^{\rm{I}}$ is
$N+1$-dimensional. The standard basis for this vector space is
\begin{eqnarray}\label{eqn;BasisCase1}
\stateA{k,l}\equiv\underbrace{\theta_{3}w_1^{\ell_1-k-1}w_2^{k}}_{\rm{Space
1}}~\underbrace{\vartheta_{3}v_1^{\ell_2-l-1}v_2^{l}}_{\rm{Space
2}},
\end{eqnarray}
for all $k+l=N$. These indeed give $N+1$ different vectors.

\subsection*{Case II, $K^{\mathrm{II}}=2 N+1, K^{\mathrm{III}}=N+1$.}

For fixed $N$, the dimension of this vector space is $4N+2$. The
standard basis is
\begin{eqnarray}\label{eqn;BasisCase2}
\stateB{k,l}_1&\equiv& \underbrace{\theta_{3}w_1^{\ell_1-k-1}w_2^{k}}~\underbrace{v_1^{\ell_2-l}v_2^{l}},\nonumber\\
\stateB{k,l}_2&\equiv&\underbrace{w_1^{\ell_1-k}w_2^{k}}~\underbrace{\vartheta_{3}v_1^{\ell_2-l-1}v_2^{l}},\\
\stateB{k,l}_3&\equiv&\underbrace{\theta_{3}w_1^{\ell_1-k-1}w_2^{k}}~\underbrace{\vartheta_{3}\vartheta_{4}v_1^{\ell_2-l-1}v_2^{l-1}},\nonumber\\
\stateB{k,l}_4&\equiv&\underbrace{\theta_{3}\theta_{4}w_1^{\ell_1-k-1}w_2^{k-1}}~\underbrace{\vartheta_{3}v_1^{\ell_2-l-1}v_2^{l}},\nonumber
\end{eqnarray}
where $k+l=N$. As a lighter notation, we will from now on, with no
risk of confusion, omit indicating ``Space 1" and ``Space 2" under
the curly brackets. The ranges of $k,l$ are clear from the
explicit expressions and it is easily seen that we get $4N+2$
states.

\subsection*{Case III: $K^{\rm{II}}=2N, K^{\rm{III}}=N$}

For fixed $N=k+l$, the dimension of this vector space is $6N$. The
standard basis is
\begin{eqnarray}\label{eqn;BasisCase3}
\stateC{k,l}_1&\equiv&\underbrace{w_1^{\ell_1-k}w_2^{k}}~\underbrace{v_1^{\ell_2-l}v_2^{l}},\nonumber\\
\stateC{k,l}_2&\equiv&\underbrace{w_1^{\ell_1-k}w_2^{k}}~\underbrace{\vartheta_{3}\vartheta_{4}v_1^{\ell_2-l-1}v_2^{l-1}},\nonumber\\
\stateC{k,l}_3&\equiv&\underbrace{\theta_{3}\theta_{4}w_1^{\ell_1-k-1}w_2^{k-1}}~\underbrace{v_1^{\ell_2-l}v_2^{l}},\nonumber\\
\stateC{k,l}_4&\equiv&\underbrace{\theta_{3}\theta_{4}w_1^{\ell_1-k-1}w_2^{k-1}}~\underbrace{\vartheta_{3}\vartheta_{4}v_1^{\ell_2-l-1}v_2^{l-1}},\\
\stateC{k,l}_5&\equiv&\underbrace{\theta_{3}w_1^{\ell_1-k-1}w_2^{k}}~\underbrace{\vartheta_{4}v_1^{\ell_2-l}v_2^{l-1}},\nonumber\\
\stateC{k,l}_6&\equiv&\underbrace{\theta_{4}w_1^{\ell_1-k}w_2^{k-1}}~\underbrace{\vartheta_{3}v_1^{\ell_2-l-1}v_2^{l}}\nonumber.
\end{eqnarray}
Note that our numbering slightly differs from the one used in
Table \ref{Tab;Tensor}, in the sense that $\stateC{k,l}_{5,6}$ are
rescaled in such a way that they also have $N=k+l$, instead of
$k+l+1$ as in Table \ref{Tab;Tensor}.

It is convenient to supply all these spaces with the usual inner
product\footnote{For the purpose of the following derivation, one
can convince oneself that orthogonality of these vectors is
actually sufficient.}
\begin{eqnarray}
 ~^{\rm{A}}_j\langle k,l|m,n\rangle^{\rm{A}}_i =
 \delta_{ij}\delta_{km}\delta_{ln}.
\end{eqnarray}
We also introduce the vector spaces $V^{\rm{A}}_{k,l} = {\rm
span}\{|k,l\rangle^{\rm{A}}_{i}\}$, for $\rm{A}= \rm{I,II,III}$.
Note that for $k,l\neq0$ we have $\dim V^{\rm{I}}_{kl} = 1,\dim
V^{\rm{II}}_{kl} = 4, \dim V^{\rm{II}}_{k0} = 3,\dim
V^{\rm{II}}_{00} = 2$ etc.

\begin{figure}
  \centering
  \includegraphics[scale=.65]{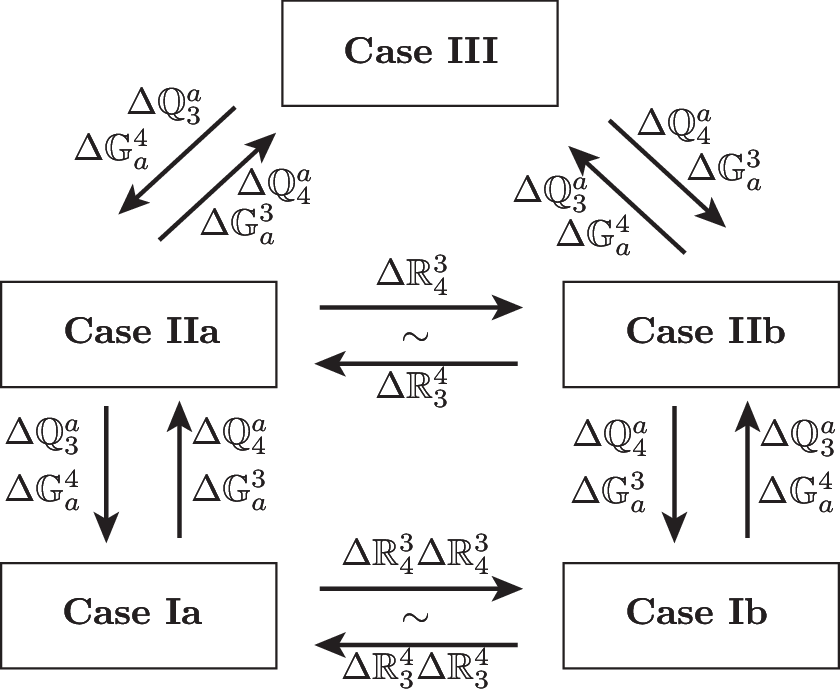}
  \caption{Schematic representation of the relations between the invariant subspaces.
  The opposite coproducts also respect the above diagram, as well as all their Yangian counterparts.}\label{Fig;Cases}
\end{figure}

The different cases are not unrelated. One can use the (opposite)
coproducts of the symmetry generators to move between the
different subspaces. In particular, the cases are distinguished by
their quantum numbers $K^{\rm{II}},K^{\rm{III}}$. Acting with
supersymmetry generators will change these numbers. Hence, these
generators provide maps between the cases. How this works is
schematically depicted in Figure \ref{Fig;Cases}.

These relations between the different cases will play an important
role in the derivation of the full S-matrix. In the next section
we will introduce two different sets of bases which allow for a
natural interpretation of the S-matrix. These bases will make use
of the full Yangian symmetry rather than just the $\alg{su}(2|2)$
as we did in this section. In this framework we can solve Case I.
Then we employ the different arrows in Figure \ref{Fig;Cases} (and
their Yangian counterparts) to relate the different S-matrices to
the Case I S-matrix.

Summarizing, we find that the S-matrix is of block-diagonal form
\begin{eqnarray}
\S=\begin{pmatrix}
  \fbox{\small{$\mathscr{X}$}} & ~ & ~ & ~ & ~ \\
  ~ & \fbox{\LARGE{$\mathscr{Y}$}} & ~ & \mbox{\Huge{$0$}} & ~ \\
  ~ & ~ & \fbox{\Huge{$\mathscr{Z}$}} & ~ & ~ \\
  ~ & \mbox{\Huge{$0$}} & ~ & \fbox{\LARGE{$\mathscr{Y}$}} & ~ \\
  ~ & ~ & ~ & ~ & \fbox{\small{$\mathscr{X}$}}
\end{pmatrix}.
\end{eqnarray}
The outer blocks scatter states from $V^{\rm{I}}$
\begin{eqnarray}
&&\mathscr{X}:V^{\rm{I}}\longrightarrow V^{\rm{I}}\\
&&\stateA{k,l}\mapsto \sum_{m=0}^{k+l}
\mathscr{X}^{k,l}_m\stateA{m,k+l-m},
\end{eqnarray}
where $\mathscr{X}^{k,l}_m$ will be given by (\ref{eqn;SCase1}).
The blocks $\mathscr{Y}$ describe the scattering of states from
$V^{\rm{II}}$
\begin{eqnarray}
&&\mathscr{Y}:V^{\rm{II}}\longrightarrow V^{\rm{II}}\\
&&\stateB{k,l}_j\mapsto \sum_{m=0}^{k+l}\sum_{j=1}^{4}
\mathscr{Y}^{k,l;j}_{m;i}\stateB{m,k+l-m}_j.
\end{eqnarray}
These S-matrix elements are given in (\ref{eqn;SCase2}). Finally,
the middle block deals with the third case
\begin{eqnarray}
&&\mathscr{Z}:V^{\rm{III}}\longrightarrow V^{\rm{III}}\\
&&\stateC{k,l}_j\mapsto \sum_{m=0}^{k+l}\sum_{j=1}^{6}
\mathscr{Z}^{k,l;j}_{m;i}\stateC{m,k+l-m}_j,
\end{eqnarray}
with $\mathscr{Z}^{k,l;j}_{m;i}$ from
(\ref{eqn;SCase3}).\smallskip

We recall that the full $\ads$ string bound state S-matrix is then
obtained by taking two copies and multiplying \emph{each one of
them} with the phase factor \cite{Arutyunov:2008zt}
\begin{eqnarray}\label{eqn;FullPhase}
S_{0}(p_{1},p_{2})&=&\left(\frac{x_{1}^{-}}{x_{1}^{+}}\right)^{\frac{\ell_2}{2}}\left(\frac{x_{2}^{+}}{x_{2}^{-}}\right)^{\frac{\ell_1}{2}}\sigma(x_{1},x_{2})\times\nonumber\\
&&\times\sqrt{G(\ell_2-\ell_1)G(\ell_2+\ell_1)}\prod_{q=1}^{\ell_1-1}G(\ell_2-\ell_1+2q),
\end{eqnarray}
where, in our conventions,
\begin{eqnarray}
\label{fattoreG}
G(Q) = \frac{u_1 - u_2 + \frac{Q}{2}}{u_1 - u_2 - \frac{Q}{2}}
\end{eqnarray}
gives the standard pole at bound-state number $Q$ (see Section
\ref{sect:EasyPoles} later on for a discussion on the physical
poles of the S-matrix). Here $u$ is given by
\begin{eqnarray}\label{eqn;defu}
u&\equiv&\frac{g}{4i}
\left(x^++\frac{1}{x^+}+x^-+\frac{1}{x^-}\right).
\end{eqnarray}

\section{Yangian Symmetry and Coproducts}\label{sect;Bases}

So far we have only used $\alg{su}(2|2)$ symmetry to study the
bound state S-matrix. This, however, is not enough to fix the
S-matrix (up to an overall phase). In particular, it was found
that one needs to impose the Yang-Baxter equation by hand to
attain this \cite{Arutyunov:2008zt}. An alternative to this method
was shown to come from Yangian symmetry \cite{deLeeuw:2008dp}.

The Yangian of $\alg{su}(2|2)$ has a Hopf-algebra
structure. In this language the invariance of the S-matrix can be
formulated as
\begin{eqnarray}\label{eqn;SymmProp}
\S~\Delta(\mathbb{J}^{A})&=&\Delta^{op}(\mathbb{J}^{A})~\S,\nonumber\\
\S~\Delta(\hat{\mathbb{J}}^{A})&=&\Delta^{op}(\hat{\mathbb{J}}^{A})~\S,
\end{eqnarray}
where $\hat{\mathbb{J}}$ stands for any generator of the Yangian.
For explicit formulae and more details we refer to Appendix
\ref{App;Yangian}. All this seems to indicate that the full
Yangian of $\alg{su}(2|2)$ should be viewed as the underlying
symmetry algebra of scattering processes. Indeed, as we will see
later on, we are able to construct any bound state S-matrix from
this algebra.

\subsection{(Opposite) coproduct basis}

Let us turn back to the invariant subspaces. In this section, we
define different bases for each case apart from the standard
basis, which is commonly used in the literature. We call them the
coproduct basis and the opposite coproduct basis. The basis
transformation between the coproduct (opposite coproduct) basis
and the standard one will be denoted by $\Lambda$ ($\Lambda^{op}$,
respectively).

These bases will be constructed by using Yangian generators to
create states out of a chosen vacuum. This is similar to
\cite{deLeeuw:2008ye} where it was used to study the Bethe Ansatz.
We define our vacuum to be
\begin{eqnarray}
|0\rangle \equiv w_1^{\ell_1}\ v_1^{\ell_2},
\end{eqnarray}
just as it is used in the coordinate Bethe Ansatz. We normalize
our S-matrix in such a way that $\mathbb{S}|0\rangle=|0\rangle$.
The (opposite) coproduct basis will consist of states created by
the (opposite) coproducts of various symmetry generators acting on
this vacuum. Clearly, the S-matrix has a natural interpretation in
these bases, and can be formulated in terms of $\Lambda$ and
$\Lambda^{op}$, as will be explained in section
\ref{sect;SmatCoProd}. We will now list the explicit formulae for
the different bases.

\subsection*{Case I, $K^{\mathrm{II}}=2 N +2, K^{\mathrm{III}}=N+2$.}

The coproduct basis is given by
\begin{eqnarray}\label{eqb;CoProdBasisCase1}
\Delta(\mathbb{Q}^{1}_{3})\Delta(\mathbb{G}_{2}^{4})\prod_{i=q+1}^{N}\Delta(\mathbb{L}^{1}_{2})
\prod_{j=1}^{q}\Delta(\hat{\mathbb{L}}^{1}_{2})|0\rangle, \qquad
q=0,1,\ldots N,
\end{eqnarray}
and the opposite coproduct basis is given by
\begin{eqnarray}
\Delta^{op}(\mathbb{Q}^{1}_{3})\Delta^{op}(\mathbb{G}_{2}^{4})\prod_{i=k+1}^{N}\Delta^{op}(\mathbb{L}^{1}_{2})
\prod_{j=1}^{k}\Delta^{op}(\hat{\mathbb{L}}^{1}_{2})|0\rangle,
\qquad k=0,1,\ldots N.
\end{eqnarray}
Each of these two bases is indeed composed of $N+1$ different
vectors. By explicitly working out the coproducts one can see that
these vectors form a basis for Case I. One could also consider an
alternative choice, like for instance
\begin{eqnarray}
\Delta(\mathbb{Q}^{1}_{3})\Delta(\hat{\mathbb{Q}}^{1}_{3})\prod_{i=k+1}^{N}\Delta(\mathbb{L}^{1}_{2})
\prod_{j=1}^{k}\Delta(\hat{\mathbb{L}}^{1}_{2})|0\rangle,
\end{eqnarray}
but these vectors are readily seen to be proportional to
(\ref{eqb;CoProdBasisCase1}).

It is also straightforwardly seen why (\ref{eqb;CoProdBasisCase1})
actually describes Case I from the point of view of the quantum
numbers $K^{\rm{II}}, K^{\rm{III}}$. The operators
$\Delta\mathbb{L}^1_2,\Delta\hat{\mathbb{L}}^1_2$ create a boson
of type $2$ out of the vacuum and the supersymmetry generators
$\Delta\mathbb{Q}^{1}_{3},\Delta\hat{\mathbb{Q}}^{1}_{3}$ create a
fermion of type $3$. Hence we find that $K^{\rm{II}} = 2\#
\mathbb{L}^1_2 + 2 \# \Delta\hat{\mathbb{L}}^1_2 + \#
\Delta\mathbb{Q}^{1}_{3} + \# \Delta\hat{\mathbb{Q}}^{1}_{3}$ and
$K^{\rm{III}} = \# \mathbb{L}^1_2 + \# \Delta\hat{\mathbb{L}}^1_2
+ \# \Delta\mathbb{Q}^{1}_{3} + \#\Delta\hat{\mathbb{Q}}^{1}_{3}$.
This indeed coincides with $K^{\mathrm{II}}=2 N +2,
K^{\mathrm{III}}=N+2$.

\subsection*{Case II, $K^{\mathrm{II}}=2 N+1, K^{\mathrm{III}}=N+1$.}

The coproduct basis is given by
\begin{eqnarray}
&&\Delta(\mathbb{Q}^{1}_{3})\prod_{i=q+1}^{N}\Delta(\mathbb{L}^{1}_{2})
\prod_{j=1}^{q}\Delta(\hat{\mathbb{L}}^{1}_{2})|0\rangle,\nonumber\\
&&\Delta(\hat{\mathbb{Q}}^{1}_{3})\prod_{i=q+1}^{N}\Delta(\mathbb{L}^{1}_{2})
\prod_{j=1}^{q}\Delta(\hat{\mathbb{L}}^{1}_{2})|0\rangle,\nonumber\\
&&\Delta(\mathbb{Q}^{1}_{3})\Delta(\hat{\mathbb{Q}}^{1}_{3})\Delta(\mathbb{Q}^{1}_{4})\prod_{i=q+1}^{N-1}\Delta(\mathbb{L}^{1}_{2})
\prod_{j=1}^{q}\Delta(\hat{\mathbb{L}}^{1}_{2})|0\rangle,\nonumber\\
&&\Delta(\mathbb{Q}^{1}_{3})\Delta(\hat{\mathbb{Q}}^{1}_{3})\Delta(\hat{\mathbb{Q}}^{1}_{4})\prod_{i=q+1}^{N-1}\Delta(\mathbb{L}^{1}_{2})
\prod_{j=1}^{q}\Delta(\hat{\mathbb{L}}^{1}_{2})|0\rangle,
\end{eqnarray}
and similar expressions hold for the opposite coproduct basis. One
can again compute $K^{\rm{II}}, K^{\rm{III}}$ for these states and
see explicitly that they describe Case II.

\subsection*{Case III, $K^{\mathrm{II}}=2 N, K^{\mathrm{III}}=N$.}

The coproduct basis is
\begin{eqnarray}
&&\prod_{i=q+1}^{N}\Delta(\mathbb{L}^{1}_{2})
\prod_{j=1}^{q}\Delta(\hat{\mathbb{L}}^{1}_{2})|0\rangle,\nonumber\\
&&\Delta(\mathbb{Q}^{1}_{3})\Delta(\mathbb{Q}^{1}_{4})\prod_{i=q+1}^{N-1}\Delta(\mathbb{L}^{1}_{2})
\prod_{j=1}^{q}\Delta(\hat{\mathbb{L}}^{1}_{2})|0\rangle,\nonumber\\
&&\Delta(\mathbb{Q}^{1}_{3})\Delta(\hat{\mathbb{Q}}^{1}_{4})\prod_{i=q+1}^{N-1}\Delta(\mathbb{L}^{1}_{2})
\prod_{j=1}^{q}\Delta(\hat{\mathbb{L}}^{1}_{2})|0\rangle,\nonumber\\
&&\Delta(\hat{\mathbb{Q}}^{1}_{3})\Delta(\mathbb{Q}^{1}_{4})\prod_{i=q+1}^{N-1}\Delta(\mathbb{L}^{1}_{2})
\prod_{j=1}^{q}\Delta(\hat{\mathbb{L}}^{1}_{2})|0\rangle,\nonumber\\
&&\Delta(\hat{\mathbb{Q}}^{1}_{3})\Delta(\hat{\mathbb{Q}}^{1}_{4})\prod_{i=q+1}^{N-1}\Delta(\mathbb{L}^{1}_{2})
\prod_{j=1}^{q}\Delta(\hat{\mathbb{L}}^{1}_{2})|0\rangle,\nonumber\\
&&\Delta(\mathbb{Q}^{1}_{3})\Delta(\mathbb{Q}^{1}_{4})\Delta(\hat{\mathbb{Q}}^{1}_{3})
\Delta(\hat{\mathbb{Q}}^{1}_{4})\prod_{i=q+1}^{N-2}\Delta(\mathbb{L}^{1}_{2})
\prod_{j=1}^{q}\Delta(\hat{\mathbb{L}}^{1}_{2})|0\rangle.\nonumber
\end{eqnarray}
These are readily seen to be $6N$ states and their quantum numbers
are of the form $K^{\mathrm{II}}=2 N, K^{\mathrm{III}}=N$.

The Yangian generators also provide maps between the different
cases. In particular, one finds that Figure \ref{Fig;Cases} also
holds for Yangian generators. One important thing to notice is the
following. Even though, for example, $\Delta\mathbb{Q}^2_{3}$ maps
Case II onto Case I, this does not automatically give a
straightforward map between the vector spaces $V_{k,l}^{\rm{A}}$.
For instance, one has
\begin{eqnarray}
&&\Delta\mathbb{Q}^2_{3}:V_{k,l}^{\rm{II}}\longrightarrow V_{k,l-1}^{\rm{I}}\oplus V_{k-1,l}^{\rm{I}},\\
&&\Delta\hat{\mathbb{Q}}^2_{3}:V_{k,l}^{\rm{II}}\longrightarrow
V_{k+1,l-1}^{\rm{I}}\oplus V_{k,l}^{\rm{I}}\oplus
V_{k-1,l+1}^{\rm{I}}.
\end{eqnarray}
These relations between the different cases will be used later on.

\subsection{S-matrix in coproduct basis}\label{sect;SmatCoProd}

The fact that the coproduct basis is well suited for computing the
S-matrix can be seen from (\ref{eqn;SymmProp}). One sees that the
S-matrix directly maps the coproduct basis onto the opposite
coproduct basis. In particular, since we normalize the S-matrix in
such a way that $\mathbb{S}|0\rangle=|0\rangle$, we see that the
S-matrix, when written as a map between these two bases, is just
the identity matrix.\smallskip

In other words, one can now get the general formula for the
S-matrix in the standard basis just by applying the appropriate
basis transformations. Let us denote the S-matrix written in the
standard basis as $\mathbb{S}$. One then finds:
\begin{eqnarray}\label{eqn;Smat}
\mathbb{S} = \Lambda^{op} \Lambda^{-1}.
\end{eqnarray}
Note that the explicit matrices $\Lambda$ and $\Lambda^{op}$
respectively, just consist of the coproduct vectors written in the
standard basis. The above discussion can be summarized in the
following commutative diagram:
\begin{eqnarray}
\begin{CD}
\{\rm{coproduct\ basis}\} @>\mathbbm{1}>>
\{\rm{opposite\ coproduct\ basis}\}\\
@V\Lambda VV @V\Lambda^{op}VV\\
\{\rm{standard\ basis}\} @>\mathbb{S}>> \{\rm{standard\ basis}\}.
\end{CD}
\end{eqnarray}
The computationally hard part is finding the explicit inverse of
$\Lambda$. For any explicit case at hand this can be done by
simple linear algebra, but the expressions become rather involved.
However, we will be able to carry out this procedure in full
generality for the S-matrix of Case I, and use this result to find
the S-matrix for all the other states. To illustrate the above
discussion, we conclude this section with an explicit example.

One of the easiest examples is that of $K^{\mathrm{II}}=1$, which
can be considered as a warm-up for Case II. The vector space
consisting of these states is only two dimensional. The standard
basis vectors are
\begin{eqnarray}
\theta_{3}w_1^{\ell_1-1}v_1^{\ell_2},\qquad
w_1^{\ell_1}\vartheta_{3}v_1^{\ell_2-1}.
\end{eqnarray}
The coproduct basis is given by
\begin{eqnarray}
\Delta(\mathbb{Q}^1_{3})|0\rangle,\qquad
\Delta(\hat{\mathbb{Q}}^1_{3})|0\rangle.
\end{eqnarray}
Explicitly, when written down in terms of the standard basis, this
gives:
\begin{eqnarray}
\Delta(\mathbb{Q}^1_{3})|0\rangle&=&a_1 \ell_1
\theta_{3}w_1^{\ell_1-1}v_1^{\ell_2} + a_2 \ell_2
w_1^{\ell_1}\vartheta_{3}v_1^{\ell_2-1} \nonumber\\
\Delta(\hat{\mathbb{Q}}^1_{3})|0\rangle&=& \frac{\ell_1\{\zeta a_2
(a_1d_2-b_2c_1)\ell_2 -i a_1
u_1\}}{2\zeta}\theta_{3}w_1^{\ell_1-1}v_1^{\ell_2}+\\
&&+\frac{\ell_2\{\zeta a_1 (b_1c_2-a_2d_1)\ell_1 -i a_2
u_2\}}{2\zeta}w_1^{\ell_1}\vartheta_{3}v_1^{\ell_2-1},\nonumber
\end{eqnarray}
or, more conveniently written in matrix form,
\begin{eqnarray}
\Lambda = \begin{pmatrix}
  a_1 \ell_1 ~ & ~ \frac{\ell_1\{\zeta a_2 (a_1d_2-b_2c_1)\ell_2 -i a_1
u_1\}}{2\zeta} \\
  a_2 \ell_2 ~ & ~ \frac{\ell_2\{\zeta a_1 (-a_2d_1+b_1c_2)\ell_1 -i a_2
u_2\}}{2\zeta}
\end{pmatrix}.
\end{eqnarray}
One obtains a similar expression for the basis transformation
concerning $\Lambda^{op}$.\smallskip

It is now an easy exercise to compute $\mathbb{S}$ for this case
via (\ref{eqn;Smat}). Doing the algebra gives the following
result:
\begin{eqnarray}\label{eqn;SmatExample}
\mathbb{S} = \begin{pmatrix}
  e^{-i\frac{p_2}{2}}\frac{x^{+}_{1}-x^{+}_{2}}{x^{+}_{1}-x^{-}_{2}} ~&~  \frac{\sqrt{\ell_1}\eta(p_{1})}{\sqrt{\ell_2}\eta(p_{2})}\frac{x^{+}_{2}-x^{-}_{2}}{x^{+}_{1}-x^{-}_{2}} \\
\frac{e^{i\frac{p_1}{2}}}{e^{i\frac{p_2}{2}}}\frac{\sqrt{\ell_2}\eta(p_{2})}{\sqrt{\ell_1}\eta(p_{1})}\frac{x^{+}_{1}-x^{-}_{1}}{x^{+}_{1}-x^{-}_{2}}
  ~&~ e^{i\frac{p_1}{2}}\frac{x^{-}_{1}-x^{-}_{2}}{x^{+}_{1}-x^{-}_{2}}
\end{pmatrix},
\end{eqnarray}
which indeed coincides with the known S-matrices
\cite{Beisert:2005tm,Arutyunov:2006yd,Arutyunov:2008zt,Bajnok:2008bm},
and also with the coefficients recently found in
\cite{deLeeuw:2008ye}.

\section{Complete Solution of Case I}

In this section, we will discuss the S-matrix of Case I . We will
use Yangian symmetry to derive its explicit form. This S-matrix
proves to be the building block out of which the S-matrices for
both Case II and Case III can be constructed\footnote{Our
procedure will somehow be reminiscent of employing highest weight
states of Yangians.}. Because of this, we will study it in detail
before moving on to the other cases. We will analyze its pole
structure and also compare this S-matrix in the semi-classical
limit to the classical $r$-matrix proposed in
\cite{Beisert:2007ty}.

\subsection{Explicit solution}

Our starting point is the coproduct basis
(\ref{eqb;CoProdBasisCase1}).  It is convenient to reorder the
products in the following way:
\begin{eqnarray}
\left\{\prod_{i=q+1}^{N}\Delta(\mathbb{L}^{1}_{2})
\prod_{j=1}^{q}\Delta(\hat{\mathbb{L}}^{1}_{2})\right\}\Delta(\mathbb{Q}^{1}_{3})\Delta(\mathbb{G}^{4}_{2})|0\rangle.
\end{eqnarray}
The action of the susy generators on the vacuum is of the form
\begin{eqnarray}
\Delta(\mathbb{Q}^{1}_{3})\Delta(\mathbb{G}^{4}_{2})|0\rangle =
(a_2c_1-a_1c_2)\ell_1\ell_2|0,0\rangle^{\rm{I}},
\end{eqnarray}
with a similar expression for the opposite version. As a matter of
fact, from this one can straightforwardly read off the action of
the S-matrix on $|0,0\rangle^{\rm{I}}$:
\begin{eqnarray}\label{eqn;Sfor00}
\mathbb{S}|0,0\rangle^{\rm{I}}&=&
\frac{\mathbb{S}\Delta(\mathbb{Q}^{1}_{3})\Delta(\mathbb{G}^{4}_{2})|0\rangle}{(a_2c_1-a_1c_2)\ell_1\ell_2}\nonumber\\
&=&\frac{\Delta^{op}(\mathbb{Q}^{1}_{3})\Delta^{op}(\mathbb{G}^{4}_{2})\mathbb{S}|0\rangle}{(a_2c_1-a_1c_2)\ell_1\ell_2}\nonumber\\
&=&\frac{a_4c_3-a_3c_4}{a_2c_1-a_1c_2}|0,0\rangle^{\rm{I}}.
\end{eqnarray}
In other words, the S-matrix multiplies $|0,0\rangle^{\rm{I}}$
by a scalar. We will denote this scalar by $D$. In terms of
$x^{\pm}$, it is given by
\begin{eqnarray}
\label{D}
D\equiv\frac{a_4c_3-a_3c_4}{a_2c_1-a_1c_2} = \frac{x_1^-
-x_2^+}{x_1^+ -x_2^-}\frac{e^{i\frac{p_1}{2}}}{e^{i\frac{p_2}{2}}}.
\end{eqnarray}
It is readily seen that this coefficient is indeed consistent with
the action of the S-matrices previously found in
\cite{Beisert:2005tm,Arutyunov:2006yd,Arutyunov:2008zt,Bajnok:2008bm}.\smallskip

One can now use the generators
$\mathbb{L}^1_2,\hat{\mathbb{L}}^1_2$ to construct a generic Case
I state $|k,l\rangle^{\rm{I}}$ from $|0,0\rangle^{\rm{I}}$, for
arbitrary $k,l$. This can be seen by considering the following
identities:
\begin{eqnarray}\label{eqn;Lplusmin}
(\mathbb{L}^1_2\otimes\mathbbm{1})(\delta
u+\Delta(\mathbb{L}^1_1))|k,l\rangle^{\rm{I}}&=&
\left\{\Delta(\hat{\mathbb{L}}^1_2)-u_2 \Delta(\mathbb{L}^1_2)+
\Delta(\mathbb{L}^1_2)\circ(\mathbb{L}^1_1\otimes
\mathbbm{1})\right\}|k,l\rangle^{\rm{I}},\\
(\mathbbm{1}\otimes\mathbb{L}^1_2)(\delta
u+\Delta(\mathbb{L}^1_1))|k,l\rangle^{\rm{I}}&=&
-\left\{\Delta(\hat{\mathbb{L}}^1_2)-u_1 \Delta(\mathbb{L}^1_2)-
\Delta(\mathbb{L}^1_2)\circ(\mathbbm{1}\otimes\mathbb{L}^1_1)\right\}|k,l\rangle^{\rm{I}},\nonumber
\end{eqnarray}
where
\begin{eqnarray}
\delta u = u_1-u_2.
\end{eqnarray}
Since $\Delta(\mathbb{L}^1_1)|k,l\rangle^{\rm{I}} =
\frac{\ell_1+\ell_2-2(k+l+1)}{2}|k,l\rangle^{\rm{I}}$, it is
obvious that the left hand side is proportional to $|k+1,l\rangle$
and $|k,l+1\rangle$ respectively. By applying these operators
inductively to $|0,0\rangle$, one finds
\begin{eqnarray}
&&\left\{\prod_{m=1}^{k}(\ell_1-m)\prod_{n=1}^{l}(\ell_2-n)\prod_{q=1}^{k+l}\left(\delta u+\frac{\ell_1+\ell_2}{2}-q\right)\right\}|k,l\rangle
=\nonumber\\
&&\qquad\left[(\mathbb{L}^1_2\otimes\mathbbm{1})(\delta
u+\Delta(\mathbb{L}^1_1))\right]^k\left[(\mathbbm{1}\otimes\mathbb{L}^1_2)(\delta
u+\Delta(\mathbb{L}^1_1))\right]^l|0, 0\rangle.
\end{eqnarray}
Then, by (\ref{eqn;Lplusmin}),
\begin{eqnarray}
&&|k,l\rangle^{\rm{I}}=\\
&&\quad \ \ \
\frac{\prod_{i=1}^k\left[\Delta(\hat{\mathbb{L}}^1_2)+\frac{\ell_1-2u_2-2i+1}{2}
\Delta(\mathbb{L}^1_2)\right]\prod_{j=1}^l\left[-\Delta(\hat{\mathbb{L}}^1_2)-\frac{1+2j-2u_1-\ell_2}{2}
\Delta(\mathbb{L}^1_2)\right]}
{\prod_{m=1}^{k}(\ell_1-m)\prod_{n=1}^{l}(\ell_2-n)\prod_{q=1}^{k+l}\left(\delta u+\frac{\ell_1+\ell_2}{2}-q\right)}|0,0\rangle^{\rm{I}}.\nonumber
\end{eqnarray}
This exactly tells us how to write a state in the standard basis
as a combination of coproducts. In other words, this explicitly
indicates how to construct $\Lambda^{-1}$. It is now
straightforward to obtain the action of the S-matrix on Case I
states from this. The symmetry properties of the S-matrix,
together with (\ref{eqn;Sfor00}), now imply
\begin{eqnarray}\label{eqn;EasyCaseSL}
&&\mathbb{S}|k,l\rangle^{\rm{I}} =\\
&& \ \
D\frac{\prod_{i=1}^k\left[\Delta^{op}(\hat{\mathbb{L}}^1_2)-\frac{2u_2-\ell_1+2i-1}{2}
\Delta^{op}(\mathbb{L}^1_2)\right]\prod_{j=1}^l\left[\frac{2u_1+\ell_2-1-2j}{2}
\Delta^{op}(\mathbb{L}^1_2)-\Delta^{op}(\hat{\mathbb{L}}^1_2)\right]}
{\prod_{m=1}^{k}(\ell_1-m)\prod_{n=1}^{l}(\ell_2-n)\prod_{q=1}^{k+l}\left(\delta
u+\frac{\ell_1+\ell_2}{2}-q\right)}|0,0\rangle^{\rm{I}}.\nonumber
\end{eqnarray}
The right hand side can be computed straightforwardly. One finds
that $\mathbb{S}|k,l\rangle^{\rm{I}}$ is of the form
\begin{eqnarray}\label{transit}
\mathbb{S}|k,l\rangle^{\rm{I}} =
\sum_{n=0}^{k+l}\mathscr{X}^{k,l}_n|n,k+l-n\rangle^{\rm{I}},
\end{eqnarray}
with
\begin{eqnarray}\label{eqn;SCase1}
\mathscr{X}^{k,l}_{n} &=&
D\frac{\prod_{i=1}^{n}(\ell_1-i)\prod_{i=1}^{k+l-n}(\ell_2-i)}{\prod_{p=1}^{k}(\ell_1-p)\prod_{p=1}^{l}(\ell_2-p)\prod_{p=1}^{k+l}(\delta u +\frac{\ell_1+\ell_2}{2}-p)}\times \\
&&\times \sum_{m=0}^{k}\left\{ {k\choose k-m }{l\choose n-m
}\prod_{p=1}^{m}\mathfrak{c}^+_p
\prod_{p=1-m}^{l-n}\mathfrak{c}^-_p
\prod_{p=1}^{k-m}\mathfrak{d}_{\frac{k-p+2}{2}}
\prod_{p=1}^{n-m}\tilde{\mathfrak{d}}_{\frac{k+l-m-p+2}{2}}\right\}.\nonumber
\end{eqnarray}
The coefficients are given by
\begin{eqnarray}
\mathfrak{c}^{\pm}_m&=&\delta u \pm \frac{\ell_1-\ell_2}{2} -m+1,\nonumber\\
\tilde{\mathfrak{c}}^{\pm}_m&=&\delta u \pm
\frac{\ell_1+\ell_2}{2} -m+1,\\
\mathfrak{d}_i&=&\ell_1+1-2i,\nonumber\\
\tilde{\mathfrak{d}}_i&=&\ell_2+1-2i.\nonumber
\end{eqnarray}
It is worthwhile to note that in the special case $l=0$ (and
similarly for $k=0$) this expression reduces considerably. For
later use, we can write it in the following way:
\begin{eqnarray}
\label{primalzero} \mathscr{X}^{k,0}_{k-n} = D{k\choose n}
\frac{\prod_{p=1}^{n}(\ell_2-p)\prod_{p=1}^{k-n}(\delta u
+\frac{\ell_1-\ell_2}{2}-p+1)}{\prod_{p=1}^{k}(\delta
u+\frac{\ell_1+\ell_2}{2}-p)}.
\end{eqnarray}
In all of the above expressions it is understood that products are
set to 1 whenever they run over negative integers, i.e.
$\prod_a^b=1$ if $b< a$, and the binomial ${x\choose y}$ is taken
to be zero if $y>x$ and if $y<0$.\smallskip

We can notice how the formula we have found bears a rational
dependence on the difference of the spectral parameters, as
typical of Yangian universal R-matrices in evaluation
representations. {\rm The following function, meromorphic in all
the parameters, coincides with  (\ref{eqn;SCase1}) in the
appropriate domain of integer values:}
\begin{eqnarray}
\label{hypergeom} \mathscr{X}^{k,l}_n &=&(-1)^{k+n} \, \pi D
\frac{\sin [(k-\ell_1) \pi ] \, \Gamma (l+1)}{\sin [\ell_1 \pi]
\sin [(k +l -\ell_2-n) \pi ] \, \Gamma (l-\ell_2+1) \Gamma
   (n+1)} \times \nonumber\\
&& \frac{\Gamma
   (n+1-\ell_1) \Gamma
   \left(l+\frac{\ell_1-\ell_2}{2}-n-\delta u
   \right) \Gamma \left(1-\frac{\ell_1+\ell_2}{2}-\delta u \right)}{\Gamma
   \left(k+l-\frac{\ell_1+\ell_2}{2}-\delta u +1\right) \Gamma \left(\frac{\ell_1-\ell_2}{2}- \delta u \right)} \times \\
&& _4\tilde{F}_3
   \left(-k,-n,\delta u
   +1-\frac{\ell_1-\ell_2}{2} ,\frac{\ell_2-\ell_1}{2}-\delta u; 1-\ell_1,\ell_2-k-l,l-n+1;1 \right),\nonumber
\end{eqnarray}
where one has defined $_4\tilde{F}_3 (x,y,z,t;r,v,w;\tau) = {_4F_3}
(x,y,z,t;r,v,w;\tau)/[\Gamma (r) \Gamma (v) \Gamma (w)]$.

Moreover, we can easily see that we are in a special situation,
since the parameters entering the hypergeometric function $_4F_3
(a_1,a_2,a_3,a_4;b_1,b_2,b_3;1)$ satisfy $\sum_{i=1}^4 a_i -
\sum_{j=1}^3 b_j=-1$. When this happens, the hypergeometric
function reduces to a $6j$-symbol, according to the following
formula (see for example \cite{shbook}):

\begin{eqnarray}
\label{6jsy}
&&_4F_3\left(a_1,a_2,a_3,a_4;b_1,b_2,b_3;1\right)=\frac{(-1)^{b_1+1}\Gamma
\left(b_2\right) \Gamma \left(b_3\right)\sqrt{\Gamma
\left(1-a_1\right)\Gamma \left(1-a_2\right)\Gamma
   \left(1-a_3\right)} }{\Gamma \left(1-b_1\right) \sqrt{\Gamma
\left(b_2-a_1\right)\Gamma \left(b_2-a_2\right)}}\times \nonumber\\
&& \qquad \frac{\sqrt{\Gamma \left(1-a_4\right)\Gamma
\left(a_1-b_1+1\right)\Gamma \left(a_2-b_1+1\right)\Gamma
\left(a_3-b_1+1\right)\Gamma \left(a_4-b_1+1\right)}}{\sqrt{\Gamma
\left(b_2-a_3\right)\Gamma
   \left(b_2-a_4\right)\Gamma \left(b_3-a_1\right)\Gamma \left(b_3-a_2\right)\Gamma \left(b_3-a_3\right)\Gamma
   \left(b_3-a_4\right)}}\times \nonumber \\
&&\qquad \, \,  \left\{\begin{array}{ccc}
 \frac{1}{2} \left(-a_1-a_4+b_3-1\right) & \frac{1}{2} \left(-a_1-a_3+b_2-1\right) & \frac{1}{2} \left(a_1+a_2-b_1-1\right) \\
 \frac{1}{2} \left(-a_2-a_3+b_3-1\right) & \frac{1}{2} \left(-a_2-a_4+b_2-1\right) & \frac{1}{2} \left(a_3+a_4-b_1-1\right)
\end{array}
\right\}.
\end{eqnarray}
By identifying the parameters we see that the relevant $6j$-symbol
\begin{eqnarray}
\left\{ \begin{array}{ccc}
 j_1 & j_2 & j_3 \\
 j_4 & j_5 & j_6
\end{array}\right\}
\end{eqnarray}
has coefficients
\begin{eqnarray}
\label{co6j}
&&j_1 = \frac{1}{2}\left(k+l-n+ \frac{\ell_1-\ell_2}{2} +\delta u\right),\nonumber\\
&&j_2 = \frac{1}{2}\left( \frac{\ell_1+\ell_2}{2} -2 - l  -
\delta u\right) ,\nonumber\\
&&j_3 = \frac{1}{2}\left( \ell_1 - 2 - k - n\right),\nonumber\\
&&j_4 = \frac{1}{2}\left( \frac{\ell_1-\ell_2}{2}- 1 + l-\delta u\right),\nonumber\\
&&j_5 = \frac{1}{2}\left( \frac{\ell_1+\ell_2}{2}-1 - k - l + n + \delta u\right),\nonumber\\
&&j_6=  \frac{1}{2}\left(\ell_2-1\right).
\end{eqnarray}
For generic values of $\delta u$, the $6j$-symbol is understood
{\rm in the same sense as in the comment above formula
(\ref{hypergeom}).}
However, one can prove that,
for values of $\delta u$ corresponding to the physical poles (see
also the discussion below), the entries of the $6j$-symbol are
indeed half-integer, as one may expect from the fusion rules of
$\alg{su}(2)$ representations.

In the special case $l=0$ (a similar argument would hold for
$k=0$), we can go back to expression (\ref{eqn;SCase1}), and see
that it can be casted in the following form:
\begin{eqnarray}
\label{dopolzero} \mathscr{X}^{k,0}_{k-n} = D \frac{\Gamma (k+1)
\Gamma (1+n-\ell_2) \Gamma \left(1-\frac{\ell_1+\ell_2}{2}-\delta
u \right) \Gamma
   \left(k+\frac{\ell_2}{2}-\frac{\ell_1}{2}-n-\delta u \right)}{\Gamma (1-\ell_2) \Gamma (k-n+1) \Gamma (n+1) \, \Gamma
   \left(k-\frac{\ell_1+\ell_2}{2}-\delta u +1\right) \Gamma
   \left(\frac{\ell_2-\ell_1}{2}-\delta u \right)}.\nonumber
\end{eqnarray}

\subsection{Poles}\label{sect:EasyPoles}
Next, we will analyze the pole structure of formula
(\ref{eqn;SCase1}). For simplicity, we will restrict to the
special case of expression (\ref{primalzero}) in this section. The
general case will then be analyzed in Appendix
\ref{sect;AppPoles}.

For the remainder of this section, we rescale the coupling
constant and the spectral parameters according to $g \rightarrow
g\sqrt{2}$, $u \rightarrow \frac{g}{i\sqrt{2}} u$, in order to
adapt our conventions to those of \cite{Chen:2006gq}. Formula
(\ref{primalzero}) becomes
\begin{eqnarray}
\mathscr{X}^{k,0}_{k-n} = D{k\choose n}
\frac{\prod_{p'=1}^{n}(\ell_2-p')\prod_{q=1}^{k-n}(\delta u  -
\frac{i}{g\sqrt{2}}[2q - \ell_1+\ell_2
-2])}{\prod_{p=1}^{k}(\delta u - \frac{i}{g\sqrt{2}}[2p -
\ell_1-\ell_2])}.
\end{eqnarray}
We can recognize, following \cite{Chen:2006gq}, the presence of
potential poles of this formula at bound-state rank $2p -\ell_1
-\ell_2$, which we now want to study. To begin with, in order for
these poles to be `physical' (namely, with a positive bound-state
rank), one must have $2p \geq \ell_1 +\ell_2$ (inclusion of the
``equal" case will not affect the result). Since $p$ is at most as
large as $k$, and, from the definition (\ref{transit}), $\ell_1
\geq k+1$, we conclude that one needs $\ell_1 \geq \ell_2 + 2$ and
$2 \ell_1 - 2 \geq 2k \geq \ell_1 + \ell_2$. If this holds, then
the physical poles occur for the values of $k \geq p \geq (\ell_1
+\ell_2)/2$. There are two possibilities, which we analyze in what
follows.
\begin{itemize}
\item {\it The numerator does not identically vanish, and does not have zeroes
for any physical values of the spectral parameters.} This
situation would leave physical poles uncancelled in the
denominator. This happens if \ $\ell_2 \geq n+1$ (by looking at
the product $\prod_{p'=1}^{n}(\ell_2-p')$), and if \ $\ell_1 -
\ell_2 \geq 2(k-n)$, with $k\geq n$ from (\ref{transit}).
Combining the two conditions, we obtain $k \leq n + [(\ell_1 -
\ell_2)/2] \leq - 1 + [(\ell_1 + \ell_2)/2]$, which contradicts
the original condition for physical poles given above. Therefore,
this situation cannot occur.
\item {\it The numerator does not identically vanish, but it has zeroes
at some physical values of the spectral parameters.} The latter
can potentially cancel some of the physical poles, and we want to
see whether few poles will be left uncancelled, or all of them
will be neutralized. In order to have zeroes at physical values,
we need $\ell_1 - \ell_2 \leq 2(k-n)-1$. This means that the
interesting zeroes will occur at positive bound-state ranks
running from $0$ up to $-k + [(\ell_1 +\ell_2)/2]$, while the
interesting poles occur at bound-state ranks running from $0$ up
to $n - k + 1 +[(\ell_1 - \ell_2)/2]$. But we see that the number
of these zeroes is always larger than the number of physical
poles. In fact, if one subtracts the two numbers, one gets $\ell_2
- n - 1$, which is bigger than (or at least equal to) zero in
order for the amplitude not to identically vanish (once again, by
looking at the product $\prod_{p'=1}^{n}(\ell_2-p')$). Therefore,
also this second situation cannot occur, and we conclude that we
cannot have physical poles in formula (\ref{primalzero}).
\end{itemize}
The result of this section and of Appendix \ref{sect;AppPoles} is
consistent with the fact that in the $\alg{su}(1|1)$ sector
corresponding to Case I, one does not expect any physical
bound-state poles \cite{Chen:2006gq,Arutyunov:2007tc}. The factor
of $D$ in fact cancels the s-channel pole at bound-state rank
$\ell_1 + \ell_2$ coming from the overall scalar factor, and no
physical poles are left in this amplitude.

\subsection{The classical limit}
In this section, we want to take the classical limit
\cite{Torrielli:2007mc} of the S-matrix for scattering of
arbitrary states belonging to Case I, and show that it is
reproduced by the universal formula given in
\cite{Beisert:2007ty}. For general transition amplitudes, though
only in the situation where one has at most two bound-state
components, this test has already been performed in
\cite{deLeeuw:2008dp}.

When looking at formula (\ref{eqn;SCase1}), one can see that,
besides expanding the factor $D$, one needs to expand the
remaining expression, depending only on the difference $\delta u$
of the spectral parameters, for large values of $\delta u$. In the
classical (near BMN) limit, in fact, the coupling $g$ goes to
infinity\footnote{The momentum goes then to zero as $p \sim \,
1/g$.}. With our conventions (\ref{eqn;defu}), the spectral
parameter $u$ grows linearly with $g$ in this regime, while both
parameters $x^\pm$, expressed as
\begin{eqnarray}
x^{\pm}_{i}& =&
x_{i}\left(\sqrt{1-\frac{(\ell_i/g)^{2}}{(x_{i}-\frac{1}{x_{i}})^{2}}}\pm
\frac{i \ell_i/g}{x_{i}-\frac{1}{x_{i}}}\right),
\end{eqnarray}
tend to their common classical value $x$ \cite{Arutyunov:2006iu}.

The relevant terms to the classical limit of (\ref{eqn;SCase1})
are given by the following expansion:

\begin{eqnarray}\label{eqn;SCase3class}
\mathscr{X}^{k,l}_n &\sim&
(1+D_{cl})\frac{\bigg( 1 - \frac{1}{\delta u}\sum_{p=1}^{k+l} (\frac{\ell_1+\ell_2}{2}-p)\bigg) \, \prod_{i=1}^{n}(\ell_1-i)\prod_{i=1}^{k+l-n}(\ell_2-i)}{\prod_{p=1}^{k}(\ell_1-p)\prod_{p=1}^{l}(\ell_2-p)}\times \nonumber \\
&&\times \sum_{m=0}^{k}\Bigg\{ \bigg( 1+ \frac{1}{\delta u} \, \sum_{p=1}^m (\frac{\ell_1-\ell_2}{2}+1-p)+\frac{1}{\delta u} \sum_{p=1-m}^{l-n} (\frac{\ell_2-\ell_1}{2}+1-p) \bigg)\times \nonumber\\
&& \, \times {\delta u}^{2m-k-n} \, {k\choose k-m }{l\choose n-m }
\prod_{p=1}^{k-m}\mathfrak{d}_{\frac{k-p+2}{2}}
\prod_{p=1}^{n-m}\tilde{\mathfrak{d}}_{\frac{k+l-m-p+2}{2}}
\Bigg\} ,
\end{eqnarray}
where $D_{cl}$ denotes the first order in $1/g$ of $D$. Here, we
have used the fact that the binomials enforce $l\geq n-m$, in
order to obtain the power of ${\delta u}^{2m-k-n}$. Let us start
by considering non-diagonal amplitudes, namely, $n$ different from
$k$ (cfr. (\ref{transit})). In order to do that, let us first
reduce the above formula for the case $n \geq k$. In this case,
the leading piece in the above expression is given by the term in
the sum with $m=k$ (the binomials are in this case non-zero,
since, from (\ref{transit}), one has $l\geq n-k$). The amplitude
tends to

\begin{eqnarray}\label{eqn;SCase3classngeqk}
\mathscr{X}^{k,l}_n &\sim& \frac{1}{{\delta u}^{n-k}} \,
\frac{\prod_{i=1}^{n}(\ell_1-i)\prod_{i=1}^{k+l-n}(\ell_2-i)}{\prod_{p=1}^{k}(\ell_1-p)\prod_{p=1}^{l}(\ell_2-p)}
\, {l\choose n-k }
\prod_{p=1}^{n-k}\tilde{\mathfrak{d}}_{\frac{l-p+2}{2}} .
\end{eqnarray}
As one can see, in the non-diagonal case only one of these
amplitudes actually contributes to the classical limit
(corresponding to the order $1/g$ of the scattering matrix).
Namely, only the transition from a state characterized by quantum
number $k$ to one with corresponding quantum number $n=k+1$ has
the right order, the other ones being suppressed by higher powers
of $\delta u$. In this situation, the classical amplitudes reads
\begin{eqnarray}\label{eqn;SCase3classneqkp1}
\mathscr{X}^{k,l}_{k+1} &\sim& \frac{1}{{\delta u}} \, l (\ell_1 -
k - 1) .
\end{eqnarray}
We checked that this is exactly the value one gets from applying
the universal formula of \cite{Beisert:2007ty} to this amplitude.
For convenience of the reader, we report here below their
classical r-matrix.
\begin{eqnarray}\label{eqn;Rmat}
r_{12} &=& \frac{\mathcal{T}_{12}-\Sigma\otimes
\mathbb{H}-\mathbb{H}\otimes
\Sigma}{i(u_{1}-u_{2})}-\frac{\Sigma\otimes \mathbb{H} }{iu_{2}}
+\frac{\mathbb{H}\otimes \Sigma}{iu_{1}}+
\frac{i}{2}(u_{2}^{-1}-u_{1}^{-1})\mathbb{H}\otimes \mathbb{H},
\end{eqnarray}
with
\begin{eqnarray}
\mathcal{T}_{12}=2\left(\mathbb{R}^{\
\alpha}_{\beta}\otimes\mathbb{R}^{\ \beta}_{\alpha}- \mathbb{L}^{\
a}_{b}\otimes\mathbb{L}^{\ b}_{a}+ \mathbb{G}^{\
\alpha}_{a}\otimes\mathbb{Q}^{\ a}_{\alpha}- \mathbb{Q}^{\
a}_{\alpha}\otimes\mathbb{G}^{\ \alpha}_{a}\right)
\end{eqnarray}
and
\begin{eqnarray}
\Sigma =\frac{1}{2} \frac{1}{ad+bc}\left(
w_{a}\frac{\partial}{\partial w_{a}}
-\theta_{a}\frac{\partial}{\partial\theta_{\alpha}}\right).
\end{eqnarray}

In (\ref{eqn;Rmat}), we understand all generators (taken at their
classical value) as differential operators, acting on the
appropriate monomials corresponding to Case I states. We then
compare the result with the expression we have found above for the
first order in $1/g$ of the complete S-matrix.

Next, let us consider $k \geq n$. In this case the binomials force
the leading piece in the sum to be the one with $m=n$. This reads
(quite symmetrically w.r.t the previous case)
\begin{eqnarray}\label{eqn;SCase3classkgeqn}
\mathscr{X}^{k,l}_n &\sim& \frac{1}{{\delta u}^{k-n}} \,
\frac{\prod_{i=1}^{n}(\ell_1-i)\prod_{i=1}^{k+l-n}(\ell_2-i)}{\prod_{p=1}^{k}(\ell_1-p)\prod_{p=1}^{l}(\ell_2-p)}
\, {k\choose k-n } \prod_{p=1}^{k-n}
\mathfrak{d}_{\frac{k-p+2}{2}} .
\end{eqnarray}
Analogously, only one of the non-diagonal terms has the right
falloff to be able to contribute to the classical r-matrix, namely
the amplitude for quantum numbers $k$ to $n=k-1$. The contribution
is given by
\begin{eqnarray}\label{eqn;SCase3classneqkp2}
\mathscr{X}^{k,l}_{k-1} &\sim& \frac{1}{{\delta u}} \, k (\ell_2 -
l - 1) ,
\end{eqnarray}
and we also checked it against the classical proposal of
\cite{Beisert:2007ty}.

The diagonal part, for $n=k$, is slightly more complicated. The
leading term can be obtained by specializing to $k=n$ either of
the two formulas (\ref{eqn;SCase3classngeqk}) or
(\ref{eqn;SCase3classkgeqn}), and is easily seen to be equal to
$1$, as expected. The quantum R-matrix goes in fact to the
identity in the strict classical limit. The next to leading term
of order $1/\delta u$ contributes to the classical r-matrix, and
can be straightforwardly obtained from (\ref{eqn;SCase3class}) as
\begin{eqnarray}\label{eqn;SCase3classngeqn}
\mathscr{X}^{k,l}_k -1 &\sim& D_{cl} \, + \frac{1}{{\delta u}}
\Bigg[ \sum_{p=1}^{k+l} (\frac{\ell_1+\ell_2}{2}-p) +
\sum_{p=1}^{k} (\frac{\ell_1-\ell_2}{2}+1-p)+ \sum_{p=1-k}^{l-k}
(\frac{\ell_2-\ell_1}{2}+1-p)\Bigg].\nonumber
\end{eqnarray}
This expression can be explicitly evaluated, and, after
supplementing it with the suitable overall scalar factor (cfr.
\cite{deLeeuw:2008dp}), we have checked that it precisely
corresponds to the result coming from the universal formula of
\cite{Beisert:2007ty}.

As a curiosity, we have checked that the order
${\cal{O}}(1/{g^2})$ of the Case I amplitude is completely
reproduced by half the square of the classical r-matrix. The
departure from a simple exponential series seems to reveal itself
starting from the next order ${\cal{O}}(1/{g^3})$. We plan to come
back to this issue in the future, in the light of possible
consequences for the abstract form of the universal R-matrix (cf.
for instance \cite{Khoroshkin:1994uk,Spill:2008yr}).

\section{The S-matrix for Case II}\label{sect;SmatCaseII}

In this section, we will use the S-matrix derived for Case I to
find the S-matrix for Case II.

As explained in the previous sections, $\Delta\mathbb{Q}^1_{3},
\Delta\mathbb{G}^{4}_2$ and their Yangian partners map Case II
states onto Case I states. We introduce the Case II S-matrix in
the following way
\begin{eqnarray}
\mathbb{S}\stateB{k,l}_i =
\sum_{j=1}^4\sum_{m=0}^{k+l}\mathscr{Y}^{k,l;j}_{m;i}\stateB{m,N-m}_j,
\end{eqnarray}
where again $N=k+l$. This means that the coefficients
$\mathscr{Y}^{k,l;j}_{m;i}$ actually correspond to the S-matrix
restricted to the following spaces
\begin{eqnarray}
\mathscr{Y}^{k,l;j}_{n;i}: V^{\rm{II}}_{k,l}\longrightarrow
V^{\rm{II}}_{n,N-n}.
\end{eqnarray}
Generically, both spaces are 4 dimensional, and
$\mathscr{Y}^{k,l;j}_{m;i}$ correspond to the coefficients of a
$4\times4$ matrix. One might wonder what happens for $n=0,N$ since
e.g. $\stateB{0,N}_4$, strictly speaking, does not exist. However,
it turns out that these states are always multiplied by 0. Hence,
the $4\times4$ matrix actually contains the non-generic case in
which $n=0,N$. This will be explained later on in Section
\ref{sect;Reduction} and we will continue with deriving the
generic $4\times4$ matrix.
\smallskip

By now considering the action of $\Delta^{op}\mathbb{Q}^1_{3}$, we
can relate the Case II S-matrix to (\ref{eqn;SCase1}). It is
easily checked that
\begin{eqnarray}
\Delta\mathbb{Q}^1_{3}\stateB{k,l}_j = Q_j(k,l)\stateA{k,l},
\end{eqnarray}
with
\begin{eqnarray}
\begin{array}{lll}
  Q_1(k,l) = a_2(l-\ell_2), &\qquad & Q_2(k,l) = a_1(\ell_1-k), \\
  Q_3(k,l) = b_2, &\qquad & Q_4(k,l) = -b_1.
\end{array}
\end{eqnarray}
Similar expressions are of course obtained for
$\Delta^{op}\mathbb{Q}^1_{3},\Delta^{op}\mathbb{G}^{4}_2,\Delta\mathbb{G}^{4}_2$.
We can now apply our general strategy in the following fashion:
\begin{eqnarray}\label{eqn;LHSderivationCase2}
\costateA{n,N-n}|~\Delta^{op}\mathbb{Q}^1_{3}\mathbb{S}~\stateB{k,l}_i
&=&\sum_{j=1}^4\sum_{m=0}^{k+l}\mathscr{Y}^{k,l;j}_{m;i}\
\costateA{n,N-n}
|~\Delta^{op}\mathbb{Q}^1_{3}~\stateB{m,N-m}_j\nonumber\\
&=&\sum_{j=1}^4\sum_{m=0}^{k+l}\mathscr{Y}^{k,l;j}_{m;i}Q^{op}_j(m,N-m)\
\costateA{n,N-n}\stateA{m,N-m}\nonumber\\
&=&\sum_{j=1}^4\mathscr{Y}^{k,l;j}_{n;i}Q^{op}_j(n,N-n).
\end{eqnarray}
On the other hand, we can use the symmetry properties of the
S-matrix to obtain
\begin{eqnarray}\label{eqn;RHSderivationCase2}
\costateA{n,N-n}|~\Delta^{op}\mathbb{Q}^1_{3}\mathbb{S}~\stateB{k,l}_i
&=&\costateA{n,N-n}|~\mathbb{S}\Delta\mathbb{Q}^1_{3}~\stateB{k,l}_i\nonumber\\
&=&Q_{i}(k,l)\costateA{n,N-n}|~\mathbb{S}~\stateB{k,l}\nonumber\\
&=&Q_{i}(k,l)\sum_{m=0}^{N}\mathscr{X}^{k,l}_{m}\ \costateA{n,N-n}\stateA{m,N-m}\nonumber\\
&=&Q_{i}(k,l)\mathscr{X}^{k,l}_{n}.
\end{eqnarray}
Clearly, this gives us four linear equations relating {the two}
S-matrices. A similar computation can be worked out using
$\Delta^{op}\mathbb{G}_2^{4}$, giving four additional equations.
We can cast the above formulae in a convenient matrix form:
\begin{eqnarray}\label{eqn;SmatCase2easypart}
\begin{pmatrix}
  \scriptstyle{a_{4}(N-n-\ell_2)} & \scriptstyle{a_{3}(\ell_1-n)} & \scriptstyle{b_{4}} & \scriptstyle{-b_{3}} \\
  \scriptstyle{c_{4}(N-n-\ell_2)} & \scriptstyle{c_{3}(\ell_1-n)} & \scriptstyle{d_{4}} & \scriptstyle{-d_{3}} \\
\end{pmatrix}
\mathscr{Y}^{k,l}_n = \mathscr{X}^{k,l}_n
\begin{pmatrix}
  \scriptstyle{a_{2}(l-\ell_2)} & \scriptstyle{a_{1}(\ell_1-k)} & \scriptstyle{b_{2}} & \scriptstyle{-b_{1}} \\
  \scriptstyle{c_{2}(l-\ell_2)} & \scriptstyle{c_{1}(\ell_1-k)} & \scriptstyle{d_{2}} & \scriptstyle{-d_{1}} \\
\end{pmatrix},
\end{eqnarray}
with
\begin{eqnarray}
\mathscr{Y}^{k,l}_n &\equiv& \begin{pmatrix}
  \scriptstyle{\mathscr{Y}^{k,l;1}_{n;1}} & \scriptstyle{\mathscr{Y}^{k,l;1}_{n;2}} & \scriptstyle{\mathscr{Y}^{k,l;1}_{n;3}} & \scriptstyle{\mathscr{Y}^{k,l;1}_{n;4}}  \\
  \scriptstyle{\mathscr{Y}^{k,l;2}_{n;1}} & \scriptstyle{\mathscr{Y}^{k,l;2}_{n;2}} & \scriptstyle{\mathscr{Y}^{k,l;2}_{n;3}} & \scriptstyle{\mathscr{Y}^{k,l;2}_{n;4}}  \\
  \scriptstyle{\mathscr{Y}^{k,l;3}_{n;1}} & \scriptstyle{\mathscr{Y}^{k,l;3}_{n;2}} & \scriptstyle{\mathscr{Y}^{k,l;3}_{n;3}} & \scriptstyle{\mathscr{Y}^{k,l;3}_{n;4}}  \\
  \scriptstyle{\mathscr{Y}^{k,l;4}_{n;1}} & \scriptstyle{\mathscr{Y}^{k,l;4}_{n;2}} & \scriptstyle{\mathscr{Y}^{k,l;4}_{n;3}} & \scriptstyle{\mathscr{Y}^{k,l;4}_{n;4}}
\end{pmatrix}.
\end{eqnarray}
Written in this way, the relation to (\ref{eqn;SymmProp}) becomes
apparent. However, it is clear from the above matrix equation
that, in order to fully determine $\mathscr{Y}^{k,l}_n$ (and
therefore the full Case II S-matrix ), one needs additional
equations.

\begin{figure}
  \centering
  \includegraphics{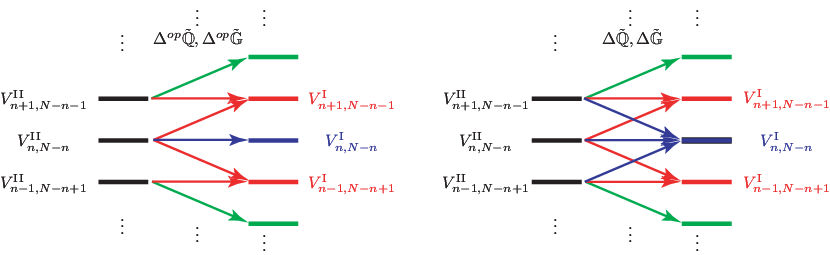}
  \caption{\label{Fig;Case2}Action of $\Delta^{op}\tilde{\mathbb{Q}},\Delta^{op}\tilde{\mathbb{G}}$ and
   $\Delta\tilde{\mathbb{Q}},\Delta\tilde{\mathbb{G}}$. They map Case II states
   (on the left) to Case I states (on the right).}
\end{figure}

These equations can be obtained via the Yangian generators.
Consider the following operators:
\begin{eqnarray}
\Delta\tilde{\mathbb{Q}}&=&\Delta(\hat{\mathbb{Q}}^1_{3})+\frac{2\Delta\hat{\mathbb{L}}^1_2\Delta(\mathbb{Q}^2_{3})}{\ell_1+\ell_2-2(N+1+\delta
u
)}-\frac{\ell_1-\ell_2+2(N-2n+u_1+u_2)}{2(\ell_1+\ell_2)-4(N+1+\delta
u )}\Delta\mathbb{L}^1_2\Delta(\mathbb{Q}^2_{3}),\nonumber\\
~\\
\Delta\tilde{\mathbb{G}}&=&\Delta(\hat{\mathbb{G}}^{4}_2)+\frac{2\Delta\hat{\mathbb{L}}^1_2\Delta(\mathbb{G}^{4}_1)}{\ell_1+\ell_2-2(N+1+\delta
u
)}+\frac{\ell_1-\ell_2+2(N-2n+u_1+u_2)}{2(\ell_1+\ell_2)-4(N+1+\delta
u )}\Delta\mathbb{L}^1_2\Delta(\mathbb{G}^{4}_1).\nonumber
\end{eqnarray}
These operators are chosen in such a way that \emph{only} states
of the form $\stateB{n,N-n}_i$ are mapped to $\stateA{n,N-n}_i$.
When we follow the same derivation as in the above, we see that
this is important in (\ref{eqn;LHSderivationCase2}) in order to be
able to pull out the matrix $\mathscr{Y}^{k,l}_n$. In fact,
$\Delta\tilde{\mathbb{Q}}$ generically maps
\begin{eqnarray}
\Delta\tilde{\mathbb{Q}}:V^{\rm{II}}_{k,l} \longrightarrow
V^{\rm{I}}_{k+1,l-1}\oplus V^{\rm{I}}_{k,l}\oplus
V^{\rm{I}}_{k-1,l+1},
\end{eqnarray}
or, more precisely, we can write
\begin{eqnarray}
&&\Delta\tilde{\mathbb{Q}}\stateB{k,l}_i =\\
&&\qquad
\tilde{Q}_i(k,l)\stateA{k,l}+\tilde{Q}^+_i(k+1,l-1)\stateA{k+1,l-1}+\tilde{Q}^-_i(k-1,l+1)\stateA{k-1,l+1}.\nonumber
\end{eqnarray}
This clearly means that, if one follows
(\ref{eqn;LHSderivationCase2}), one obtains
\begin{eqnarray}
&&\costateA{n,N-n}|~\Delta^{op}\tilde{\mathbb{Q}}\mathbb{S}~\stateB{k,l}_i
=\\
&&\qquad\sum_{j=1}^4\mathscr{Y}^{k,l;j}_{n;i}\tilde{Q}^{op}_j(n,N-n)+\mathscr{Y}^{k,l;j}_{n+1;i}\tilde{Q}^{op,+}_j(n,N-n)+\mathscr{Y}^{k,l;j}_{n-1;i}\tilde{Q}^{op,-}_j(n,N-n).\nonumber
\end{eqnarray}
However, the specific choice we made for
$\Delta\tilde{\mathbb{Q}}$ means that
$\tilde{Q}^{op,+}_j(n,N-n)=\tilde{Q}^{op,-}_j(n,N-n)=0$. In other
words, we can again pull out the matrix factor
$\mathscr{Y}^{k,l}_n$ on the left hand side of the final equation.
Since this is specifically tuned to work for the opposite
coproducts, the right hand side of the equation will not have this
property, and $\tilde{Q}^{\pm}$ will contribute there. This is
exemplified in Figure \ref{Fig;Case2}.

For compactness, let us define $M\equiv N-2n~ (= N-n-n)$. By
combining all the equations one is lead to the following matrix
equation:
\begin{eqnarray*}
&&
\begin{pmatrix}
  \scriptstyle{a_4} & \scriptstyle{a_3} & \scriptstyle{0} & \scriptstyle{0} \\
  \scriptstyle{c_4} & \scriptstyle{c_3} & \scriptstyle{0} & \scriptstyle{0} \\
  \scriptstyle{0} & \scriptstyle{0} & \scriptstyle{a_4} & \scriptstyle{a_3} \\
  \scriptstyle{0} & \scriptstyle{0} & \scriptstyle{c_4} & \scriptstyle{c_3}
\end{pmatrix}
A~\mathscr{Y}^{k,l}_n =
\begin{pmatrix}
  \scriptstyle{a_2} & \scriptstyle{a_1} & \scriptstyle{0} & \scriptstyle{0} \\
  \scriptstyle{c_2} & \scriptstyle{c_1} & \scriptstyle{0} & \scriptstyle{0} \\
  \scriptstyle{0} & \scriptstyle{0} & \scriptstyle{a_2} & \scriptstyle{a_1} \\
  \scriptstyle{0} & \scriptstyle{0} & \scriptstyle{c_2} & \scriptstyle{c_1}
\end{pmatrix}
\left\{B^+\mathscr{X}^{k+1,l-1}_n + B^-\mathscr{X}^{k-1,l+1}_n +B
\mathscr{X}^{k,l}_n \right\},~~
\end{eqnarray*}
where the matrix on the left hand side is given by
\begin{eqnarray}
A&=&
\begin{pmatrix}
  \scriptstyle{N-n-\ell_2} & \scriptstyle{0} & \scriptstyle{\frac{\mathscr{I}_{34}}{\mathscr{Q}_{34}}} & \scriptstyle{\frac{1}{\mathscr{Q}_{43}}} \\
  \scriptstyle{0} & \scriptstyle{\ell_1-n} & \scriptstyle{\frac{1}{\mathscr{Q}_{43}}} & \scriptstyle{\frac{\mathscr{I}_{43}}{\mathscr{Q}_{34}}} \\
  \scriptstyle{(N-n-\ell_2)(M-\delta u)} & \scriptstyle{(n-\ell_1)\ell_2\mathscr{I}_{34}} & \scriptstyle{\frac{(\delta u-M+\ell_2)\mathscr{I}_{34}}{\mathscr{Q}_{43}}} & \scriptstyle{\frac{\delta u+M+\ell_1-\ell_2\mathscr{Q}_{34}\overline{\mathscr{Q}}_{34}}{\mathscr{Q}_{43}}} \\
  \scriptstyle{(N-n-\ell_2)(\ell_1\mathscr{I}_{43})} & \scriptstyle{(\ell_1-n)(\delta u+M)} & \scriptstyle{\frac{M-\delta u-\ell_2+\ell_1\mathscr{Q}_{34}\overline{\mathscr{Q}}_{34}}{\mathscr{Q}_{43}}} & \scriptstyle{\frac{(\delta u+M+\ell_1)\mathscr{I}_{43}}{\mathscr{Q}_{34}}}
\end{pmatrix}
\end{eqnarray}
and the matrices on the right hand side by
\begin{eqnarray}
&&B^+=
\frac{2(\ell_1-k-1)\mathfrak{c}^-_{l-n}}{\tilde{\mathfrak{c}}^-_{-N}}
\left(
\begin{array}{cccc}
 \scriptstyle{0} & \scriptstyle{0} & \scriptstyle{0} & \scriptstyle{0} \\
 \scriptstyle{0} & \scriptstyle{0} & \scriptstyle{0} & \scriptstyle{0} \\
 \scriptstyle{l} & \scriptstyle{0} & \frac{\mathscr{I}_{12}}{\mathscr{Q}_{12}} & \scriptstyle{0} \\
 \scriptstyle{0} & \scriptstyle{0} & \frac{1}{\mathscr{Q}_{21}} & \scriptstyle{0}
\end{array}
\right),\qquad B^-=
\frac{2(\ell_2-l-1)\mathfrak{c}^+_{n-l}}{\tilde{\mathfrak{c}}^-_{-N}}
\left(
\begin{array}{cccc}
 \scriptstyle{0} & \scriptstyle{0} & \scriptstyle{0} & \scriptstyle{0} \\
 \scriptstyle{0} & \scriptstyle{0} & \scriptstyle{0} & \scriptstyle{0} \\
 \scriptstyle{0} & \scriptstyle{0} & \scriptstyle{0} & \frac{1}{\mathscr{Q}_{12}} \\
 \scriptstyle{0} & \scriptstyle{k} & \scriptstyle{0} & \frac{\mathscr{I}_{21}}{\mathscr{Q}_{21}}
\end{array}
\right)\nonumber\\
&&B=
\begin{pmatrix}
  \scriptstyle{l-\ell_2} & \scriptstyle{0} & \scriptstyle{\frac{\mathscr{I}_{12}}{\mathscr{Q}_{12}}} & \scriptstyle{\frac{1}{\mathscr{Q}_{21}}} \\
  \scriptstyle{0} & \scriptstyle{\ell_1-k} & \scriptstyle{\frac{1}{\mathscr{Q}_{21}}} & \scriptstyle{\frac{\mathscr{I}_{21}}{\mathscr{Q}_{12}}} \\
  \scriptstyle{(l-\ell_2)(N-\delta u)} & \scriptstyle{(\ell_1-k)\ell_2\mathscr{I}_{12}} & \scriptstyle{\frac{(N-\delta u-\ell_2)\mathscr{I}_{12}}{\mathscr{Q}_{12}}} & \scriptstyle{\frac{N-\delta u-\ell_1-\ell_2\mathscr{Q}_{12}\overline{\mathscr{Q}}_{12}}{\mathscr{Q}_{12}}} \\
  \scriptstyle{(\ell_2-l)(\ell_1\mathscr{I}_{21})} & \scriptstyle{(\ell_1-k)(\delta u-N)} & \scriptstyle{\frac{\delta u-N+\ell_1\mathscr{Q}_{12}\overline{\mathscr{Q}}_{12}+\ell_2}{\mathscr{Q}_{12}}} & \scriptstyle{\frac{(\delta u-N+\ell_1)\mathscr{I}_{21}}{\mathscr{Q}_{12}}}
\end{pmatrix}\nonumber\\
&&\qquad -2
\begin{pmatrix}
  \scriptstyle{0} & \scriptstyle{0} & \scriptstyle{0} & \scriptstyle{0} \\
  \scriptstyle{0} & \scriptstyle{0} & \scriptstyle{0} & \scriptstyle{0} \\
  \scriptstyle{\frac{l(1+n+k-\ell_1)(l-\ell_2)}{\tilde{\mathfrak{c}}^-_{-N}}} & \scriptstyle{0} & \scriptstyle{\frac{(l-\ell_2)(1+n+k-\ell_1)\mathscr{I}_{12}}{\tilde{\mathfrak{c}}^-_{-N}\mathscr{Q}_{12}}} & \scriptstyle{\frac{(\ell_1-k)(1+N-n+l-\ell_2)}{\tilde{\mathfrak{c}}^-_{-N}\mathscr{Q}_{21}}} \\
  \scriptstyle{0} & \scriptstyle{\frac{k(1+N-n+l-\ell_2)(k-\ell_1)}{\tilde{\mathfrak{c}}^-_{-N}}} & \scriptstyle{\frac{(l-\ell_2)(1+n+k-\ell_1)}{\tilde{\mathfrak{c}}^-_{-N}\mathscr{Q}_{21}}} & \scriptstyle{\frac{(\ell_1-k)(1+N-n+l-\ell_2)\mathscr{I}_{21}}{\tilde{\mathfrak{c}}^-_{-N}\mathscr{Q}_{12}}}
\end{pmatrix}\nonumber,
\end{eqnarray}
where we defined
\begin{eqnarray}
\mathscr{Q}_{ij}&=&a_i c_j-a_j c_i,\nonumber\\
\overline{\mathscr{Q}}_{ij}&=&b_i d_j-d_i b_j,\\
\mathscr{I}_{ij}&=&a_i d_j-b_j c_i\nonumber.
\end{eqnarray}
These coefficients satisfy the following identity
\begin{eqnarray}
\mathscr{Q}_{ij}\overline{\mathscr{Q}_{ij}} =
1-\mathscr{I}_{ij}\mathscr{I}_{ji}.
\end{eqnarray}
Notice the similarities between the matrices $A,B$ and $B^+,B^-$.
From this, it is now straightforward to extract
$\mathscr{Y}^{k,l}_n$ by simple linear algebra. For completeness,
we will explicitly report this inverse matrix. Define
\begin{eqnarray}
\mathfrak{da}\equiv \det A\frac{\mathscr{Q}_{34}^2}{(n-\ell_1)(N-n-\ell_2)} &=& -4\delta
u^2+(\ell_1-\ell_2)^2+4\ell_1\ell_2\mathscr{I}_{34}\mathscr{I}_{43}\nonumber\\
&=&-4\mathfrak{c}^+_1\mathfrak{c}^-_1+4\ell_1\ell_2\mathscr{I}_{34}\mathscr{I}_{43},
\end{eqnarray}
then
\begin{eqnarray}
&&A^{-1} =\frac{2}{\mathfrak{da}}
\begin{pmatrix}
  \scriptstyle{\frac{1}{N-n-\ell_2}} & \scriptstyle{0} & \scriptstyle{0} & \scriptstyle{0} \\
  \scriptstyle{0} & \scriptstyle{\frac{1}{n-\ell_1}} & \scriptstyle{0} & \scriptstyle{0} \\
  \scriptstyle{0} & \scriptstyle{0} & \scriptstyle{\mathscr{Q}_{43}} & \scriptstyle{0} \\
  \scriptstyle{0} & \scriptstyle{0} & \scriptstyle{0} & \scriptstyle{\mathscr{Q}_{43}}
\end{pmatrix}\times\\
&&\times
\begin{pmatrix}
  \scriptstyle{\frac{\alg{da}}{4}-\left[M+\frac{\ell_1-\ell_2}{2}\right]\left[\mathfrak{c}^-_1+\ell_1\mathscr{I}_{34}\mathscr{I}_{43}\right]} & \scriptstyle{\mathscr{I}_{34}\left(\frac{\alg{da}}{4}-\left[M+\frac{\ell_1-\ell_2}{2}\right]\tilde{\mathfrak{c}}^+_1\right)} & \scriptstyle{\mathfrak{c}^-_1+\ell_1\mathscr{I}_{34}\mathscr{I}_{43}} & \scriptstyle{\tilde{\mathfrak{c}}^+_1\mathscr{I}_{34}} \\
  \scriptstyle{-\mathscr{I}_{43}\left(\frac{\alg{da}}{4}+\left[M+\frac{\ell_1-\ell_2}{2}\right]\tilde{\mathfrak{c}}^+_1\right)} & \scriptstyle{-\left[M+\frac{\ell_1-\ell_2}{2}\right]\left[\mathfrak{c}^+_1+\ell_2\mathscr{I}_{34}\mathscr{I}_{43}\right]-\frac{\alg{da}}{4}} & \scriptstyle{\tilde{\mathfrak{c}}^+_1\mathscr{I}_{43}} & \scriptstyle{\mathfrak{c}^+_1+\ell_2\mathscr{I}_{34}\mathscr{I}_{43}} \\
  \scriptstyle{-\ell_1\left[M+\frac{\ell_1-\ell_2}{2}\right]\mathscr{I}_{43}} &  \scriptstyle{\frac{\alg{da}}{4}-\mathfrak{c}^+_1\left[M+\frac{\ell_1-\ell_2}{2}\right]} & \scriptstyle{\ell_1\mathscr{I}_{43}} & \scriptstyle{\mathfrak{c}^+_1} \\
  \scriptstyle{\frac{\alg{da}}{4}+\mathfrak{c}^-_1\left[M+\frac{\ell_1-\ell_2}{2}\right]} & \scriptstyle{\ell_2\left[M+\frac{\ell_1-\ell_2}{2}\right]\mathscr{I}_{34}} & \scriptstyle{-\mathfrak{c}^-_1} & \scriptstyle{-\ell_2\mathscr{I}_{34}}
\end{pmatrix}\nonumber
\end{eqnarray}
and
\begin{eqnarray}\label{eqn;SCase2}
\mathscr{Y}^{k,l}_n = A^{-1}
\begin{pmatrix}
  \scriptstyle{\frac{\mathscr{Q}_{32}}{\mathscr{Q}_{34}}} & \scriptstyle{\frac{\mathscr{Q}_{31}}{\mathscr{Q}_{34}}} & \scriptstyle{0} & \scriptstyle{0} \\
  \scriptstyle{\frac{\mathscr{Q}_{42}}{\mathscr{Q}_{43}}} & \scriptstyle{\frac{\mathscr{Q}_{41}}{\mathscr{Q}_{43}}} & \scriptstyle{0} & \scriptstyle{0} \\
  \scriptstyle{0} & \scriptstyle{0} & \scriptstyle{\frac{\mathscr{Q}_{32}}{\mathscr{Q}_{34}}} & \scriptstyle{\frac{\mathscr{Q}_{31}}{\mathscr{Q}_{34}}} \\
  \scriptstyle{0} & \scriptstyle{0} & \scriptstyle{\frac{\mathscr{Q}_{42}}{\mathscr{Q}_{43}}} & \scriptstyle{\frac{\mathscr{Q}_{41}}{\mathscr{Q}_{43}}}
\end{pmatrix}
\left\{\mathscr{X}^{k+1,l-1}_n B^+ + \mathscr{X}^{k-1,l+1}_n B^- +
\mathscr{X}^{k,l}_n B \right\}.
\end{eqnarray}
Note that the final result for $\mathscr{Y}^{k,l}_n$ purely
depends on the spectral parameters through their difference
$\delta u$, and the representation parameters only appear in the
combinations $\mathscr{Q}_{ij},\mathcal{H}_{ij}$ (modulo perhaps
an overall factor), the rest being taken care of by combinatorial
factors involving the integer bound-state components.

\section{Complete Solution of Case III}

We will perform here a similar construction as done in the
previous section, in order to solve Case III in terms of Case II.
Let us first set few additional notations. We introduce the
S-matrix at this level in the following way:
\begin{eqnarray}
\mathbb{S}\stateC{k,l}_i \equiv
\sum_{m=0}^{k+l}\sum_{j=1}^{6}\mathscr{Z}^{k,l;j}_{m;i}\stateC{m,k+l-m}_j.
\end{eqnarray}
It is clear that one can repeat a very similar derivation as
performed in (\ref{eqn;LHSderivationCase2}) and
(\ref{eqn;RHSderivationCase2}), where, instead of $\mathscr{X}$,
one has to think of having $\mathscr{Y}$ (and indices running over
the appropriate domains). This time, moreover, one considers the
action of $\Delta\mathbb{Q}^1_{3},\Delta\mathbb{G}_2^{4}$. The
result is now the following matrix equations:
\begin{eqnarray}
&&\begin{pmatrix}
  \scriptstyle{(n-\ell_1)a_3} & \scriptstyle{0} & \scriptstyle{b_3} & \scriptstyle{0} & \scriptstyle{b_4} & \scriptstyle{0} \\
  \scriptstyle{(N-n-\ell_2)a_4} & \scriptstyle{b_4} & \scriptstyle{0} & \scriptstyle{0} & \scriptstyle{0} & \scriptstyle{-b_3} \\
  \scriptstyle{0} & \scriptstyle{(n-\ell_1)a_3} & \scriptstyle{0} & \scriptstyle{b_3} & \scriptstyle{(N-n-\ell_2)a_4} & \scriptstyle{0} \\
  \scriptstyle{0} & \scriptstyle{0} & \scriptstyle{(N-n-\ell_2)a_4} & \scriptstyle{b_4} & \scriptstyle{0} &
  \scriptstyle{(n-\ell_1)a_3}
\end{pmatrix}\mathscr{Z}^{k,l}_n
=\\
&& \qquad \mathscr{Y}^{k,l}_n
\begin{pmatrix}
  \scriptstyle{(k-\ell_1)a_1} & \scriptstyle{0} & \scriptstyle{b_1} & \scriptstyle{0} & \scriptstyle{b_2} & \scriptstyle{0} \\
  \scriptstyle{(l-\ell_2)a_2} & \scriptstyle{b_2} & \scriptstyle{0} & \scriptstyle{0} & \scriptstyle{0} & \scriptstyle{-b_1} \\
  \scriptstyle{0} & \scriptstyle{(k-\ell_1)a_1} & \scriptstyle{0} & \scriptstyle{b_1} & \scriptstyle{(l-\ell_2)a_2} & \scriptstyle{0} \\
  \scriptstyle{0} & \scriptstyle{0} & \scriptstyle{(l-\ell_2)a_2} & \scriptstyle{b_2} & \scriptstyle{0} &
  \scriptstyle{(k-\ell_1)a_1}
\end{pmatrix},
\end{eqnarray}
and
\begin{eqnarray}
&&\begin{pmatrix}
  \scriptstyle{(n-\ell_1)c_3} & \scriptstyle{0} & \scriptstyle{d_3} & \scriptstyle{0} & \scriptstyle{d_4} & \scriptstyle{0} \\
  \scriptstyle{(N-n-\ell_2)c_4} & \scriptstyle{d_4} & \scriptstyle{0} & \scriptstyle{0} & \scriptstyle{0} & \scriptstyle{-d_3} \\
  \scriptstyle{0} & \scriptstyle{(n-\ell_1)c_3} & \scriptstyle{0} & \scriptstyle{d_3} & \scriptstyle{(N-n-\ell_2)d_4} & \scriptstyle{0} \\
  \scriptstyle{0} & \scriptstyle{0} & \scriptstyle{(N-n-\ell_2)c_4} & \scriptstyle{d_4} & \scriptstyle{0} &
  \scriptstyle{(n-\ell_1)c_3}
\end{pmatrix}\mathscr{Z}^{k,l}_n
=\\
&& \qquad\qquad\qquad\qquad\qquad \mathcal{Y}^{k,l}_n
\begin{pmatrix}
  \scriptstyle{(k-\ell_1)c_1} & \scriptstyle{0} & \scriptstyle{d_1} & \scriptstyle{0} & \scriptstyle{d_2} & \scriptstyle{0} \\
  \scriptstyle{(l-\ell_2)c_2} & \scriptstyle{d_2} & \scriptstyle{0} & \scriptstyle{0} & \scriptstyle{0} & \scriptstyle{-d_1} \\
  \scriptstyle{0} & \scriptstyle{(k-\ell_1)c_1} & \scriptstyle{0} & \scriptstyle{d_1} & \scriptstyle{(l-\ell_2)c_2} & \scriptstyle{0} \\
  \scriptstyle{0} & \scriptstyle{0} & \scriptstyle{(l-\ell_2)c_2} & \scriptstyle{d_2} & \scriptstyle{0} &
  \scriptstyle{(k-\ell_1)c_1}
\end{pmatrix}\nonumber,
\end{eqnarray}
where
\begin{eqnarray}
\mathscr{Z}^{k,l}_n &\equiv& \begin{pmatrix}
  \scriptstyle{\mathscr{Z}^{k,l;1}_{n;1}} & \scriptstyle{\mathscr{Z}^{k,l;1}_{n;2}} & \scriptstyle{\mathscr{Z}^{k,l;1}_{n;3}} & \scriptstyle{\mathscr{Z}^{k,l;1}_{n;4}} & \scriptstyle{\mathscr{Z}^{k,l;1}_{n;5}} & \scriptstyle{\mathscr{Z}^{k,l;1}_{n;6}} \\
  \scriptstyle{\mathscr{Z}^{k,l;2}_{n;1}} & \scriptstyle{\mathscr{Z}^{k,l;2}_{n;2}} & \scriptstyle{\mathscr{Z}^{k,l;2}_{n;3}} & \scriptstyle{\mathscr{Z}^{k,l;2}_{n;4}} & \scriptstyle{\mathscr{Z}^{k,l;2}_{n;5}} & \scriptstyle{\mathscr{Z}^{k,l;2}_{n;6}} \\
  \scriptstyle{\mathscr{Z}^{k,l;3}_{n;1}} & \scriptstyle{\mathscr{Z}^{k,l;3}_{n;2}} & \scriptstyle{\mathscr{Z}^{k,l;3}_{n;3}} & \scriptstyle{\mathscr{Z}^{k,l;3}_{n;4}} & \scriptstyle{\mathscr{Z}^{k,l;3}_{n;5}} & \scriptstyle{\mathscr{Z}^{k,l;3}_{n;6}} \\
  \scriptstyle{\mathscr{Z}^{k,l;4}_{n;1}} & \scriptstyle{\mathscr{Z}^{k,l;4}_{n;2}} & \scriptstyle{\mathscr{Z}^{k,l;4}_{n;3}} & \scriptstyle{\mathscr{Z}^{k,l;4}_{n;4}} & \scriptstyle{\mathscr{Z}^{k,l;4}_{n;5}} & \scriptstyle{\mathscr{Z}^{k,l;4}_{n;6}} \\
  \scriptstyle{\mathscr{Z}^{k,l;5}_{n;1}} & \scriptstyle{\mathscr{Z}^{k,l;5}_{n;2}} & \scriptstyle{\mathscr{Z}^{k,l;5}_{n;3}} & \scriptstyle{\mathscr{Z}^{k,l;5}_{n;4}} & \scriptstyle{\mathscr{Z}^{k,l;5}_{n;5}} & \scriptstyle{\mathscr{Z}^{k,l;5}_{n;6}} \\
  \scriptstyle{\mathscr{Z}^{k,l;6}_{n;1}} & \scriptstyle{\mathscr{Z}^{k,l;6}_{n;2}} & \scriptstyle{\mathscr{Z}^{k,l;6}_{n;3}} & \scriptstyle{\mathscr{Z}^{k,l;6}_{n;4}} & \scriptstyle{\mathscr{Z}^{k,l;6}_{n;5}} & \scriptstyle{\mathscr{Z}^{k,l;6}_{n;6}}
\end{pmatrix}.
\end{eqnarray}
Again the relation with (\ref{eqn;SymmProp}) is apparent.

However, it is readily checked that these equations are not
independent. Hence, once again one needs additional equations, as
in the previous section it was required in order to compute
$\mathscr{Y}$. In that case, they were provided by Yangian
generators. In this case we are more fortunate and do not need the
Yangian, since one can consider the action of
$\Delta\mathbb{Q}^2_4$ and $\Delta\mathbb{G}^3_1$. It is easy to
check that, by repeating the above procedure using these
additional symmetries, one arrives this time at the following
matrix equations:
\begin{eqnarray}
&&\begin{pmatrix}
  \scriptstyle{n a_3} & \scriptstyle{0} & \scriptstyle{b_3} & \scriptstyle{0} & \scriptstyle{0} & \scriptstyle{-b_4} \\
  \scriptstyle{(N-n)a_4} & \scriptstyle{b_4} & \scriptstyle{0} & \scriptstyle{0} & \scriptstyle{b_3} & \scriptstyle{0} \\
  \scriptstyle{0} & \scriptstyle{n a_3} & \scriptstyle{0} & \scriptstyle{b_3} & \scriptstyle{0} & \scriptstyle{(N-n) a_4} \\
  \scriptstyle{0} & \scriptstyle{0} & \scriptstyle{(N-n)a_4} & \scriptstyle{b_4} & \scriptstyle{-n a_3} &
  \scriptstyle{0}
\end{pmatrix}\mathscr{Z}^{k,l}_n
= \tilde{\mathscr{Y}}^{k,l}_n
\begin{pmatrix}
  \scriptstyle{k a_1} & \scriptstyle{0} & \scriptstyle{b_1} & \scriptstyle{0} & \scriptstyle{0} & \scriptstyle{-b_2} \\
  \scriptstyle{l a_2} & \scriptstyle{b_2} & \scriptstyle{0} & \scriptstyle{0} & \scriptstyle{b_1} & \scriptstyle{0} \\
  \scriptstyle{0} & \scriptstyle{k a_1} & \scriptstyle{0} & \scriptstyle{b_1} & \scriptstyle{0} & \scriptstyle{l a_2} \\
  \scriptstyle{0} & \scriptstyle{0} & \scriptstyle{l a_2} & \scriptstyle{b_2} & \scriptstyle{-k a_1} &
  \scriptstyle{0}
\end{pmatrix}
\end{eqnarray}
and
\begin{eqnarray}
&&\begin{pmatrix}
  \scriptstyle{n c_3} & \scriptstyle{0} & \scriptstyle{d_3} & \scriptstyle{0} & \scriptstyle{0} & \scriptstyle{-d_4} \\
  \scriptstyle{(N-n)c_4} & \scriptstyle{d_4} & \scriptstyle{0} & \scriptstyle{0} & \scriptstyle{d_3} & \scriptstyle{0} \\
  \scriptstyle{0} & \scriptstyle{n c_3} & \scriptstyle{0} & \scriptstyle{d_3} & \scriptstyle{0} & \scriptstyle{(N-n) c_4} \\
  \scriptstyle{0} & \scriptstyle{0} & \scriptstyle{(N-n)c_4} & \scriptstyle{d_4} & \scriptstyle{-n c_3} &
  \scriptstyle{0}
\end{pmatrix}\mathscr{Z}^{k,l}_n
= \tilde{\mathscr{Y}}^{k,l}_n
\begin{pmatrix}
  \scriptstyle{k c_1} & \scriptstyle{0} & \scriptstyle{d_1} & \scriptstyle{0} & \scriptstyle{0} & \scriptstyle{-d_2} \\
  \scriptstyle{l c_2} & \scriptstyle{d_2} & \scriptstyle{0} & \scriptstyle{0} & \scriptstyle{d_1} & \scriptstyle{0} \\
  \scriptstyle{0} & \scriptstyle{k c_1} & \scriptstyle{0} & \scriptstyle{d_1} & \scriptstyle{0} & \scriptstyle{l c_2} \\
  \scriptstyle{0} & \scriptstyle{0} & \scriptstyle{l c_2} & \scriptstyle{d_2} & \scriptstyle{-k c_1} &
  \scriptstyle{0}
\end{pmatrix},
\end{eqnarray}
where we have defined
\begin{eqnarray}
\tilde{\mathscr{Y}}^{k,l}_n &\equiv& \begin{pmatrix}
  \mathscr{Y}^{k-1,l;1}_{n-1;1} & \mathscr{Y}^{k,l-1;1}_{n-1;2} & \mathscr{Y}^{k-1,l;1}_{n-1;3} & \mathscr{Y}^{k,l-1;1}_{n-1;4}  \\
  \mathscr{Y}^{k-1,l;2}_{n;1} & \mathscr{Y}^{k,l-1;2}_{n;2} & \mathscr{Y}^{k-1,l;2}_{n;3} & \mathscr{Y}^{k,l-1;2}_{n;4}  \\
  \mathscr{Y}^{k-1,l;3}_{n-1;1} & \mathscr{Y}^{k,l-1;3}_{n-1;2} & \mathscr{Y}^{k-1,l;3}_{n-1;3} & \mathscr{Y}^{k,l-1;3}_{n-1;4}  \\
  \mathscr{Y}^{k-1,l;4}_{n;1} & \mathscr{Y}^{k,l-1;4}_{n;2} & \mathscr{Y}^{k-1,l;4}_{n;3} & \mathscr{Y}^{k,l-1;4}_{n;4}
\end{pmatrix}.
\end{eqnarray}
Combining all of the above equations is sufficient in order to
solve for $\mathscr{Z}$. To be more precise, one can write the
equation for $\mathscr{Z}^{k,l}_{n}$ in the following way:
\begin{eqnarray}
&&\left(
\begin{array}{cccccc}
 \scriptstyle{(n-\ell_1)\mathscr{Q}_{43}} & \scriptstyle{0} & \scriptstyle{\mathscr{I}_{43}} & \scriptstyle{0} & \scriptstyle{1} & \scriptstyle{0} \\
 \scriptstyle{0} & \scriptstyle{1} & \scriptstyle{0} & \scriptstyle{0} & \scriptstyle{0} & \scriptstyle{-\mathscr{I}_{43}} \\
 \scriptstyle{0} & \scriptstyle{(n-\ell_1)\mathscr{Q}_{43}} & \scriptstyle{0} & \scriptstyle{\mathscr{I}_{43}} & \scriptstyle{0} & \scriptstyle{0} \\
 \scriptstyle{-n \mathscr{Q}_{43}} & \scriptstyle{0} & \scriptstyle{-\mathscr{I}_{43}} & \scriptstyle{0} & \scriptstyle{0} & \scriptstyle{1} \\
 \scriptstyle{0} & \scriptstyle{-1} & \scriptstyle{0} & \scriptstyle{0} & \scriptstyle{-\mathscr{I}_{43}} & \scriptstyle{0} \\
 \scriptstyle{0} & \scriptstyle{-n \mathscr{Q}_{43}} & \scriptstyle{0} & \scriptstyle{-\mathscr{I}_{43}} & \scriptstyle{0} & \scriptstyle{0}
\end{array}
\right) \mathscr{Z}^{k,l}_n =\\
&&\qquad\qquad\qquad \check{\mathscr{S}}^{k,l}_n \left(
\begin{array}{cccccc}
  \scriptstyle{(\ell_1-k ) \mathscr{Q}_{14}} & \scriptstyle{0} & \scriptstyle{\mathscr{I}_{41}} & \scriptstyle{0} & \scriptstyle{\mathscr{I}_{42}} & \scriptstyle{0} \\
  \scriptstyle{(l-\ell_2 ) \mathscr{Q}_{42}} & \scriptstyle{\mathscr{I}_{42}} & \scriptstyle{0} & \scriptstyle{0} & \scriptstyle{0} & \scriptstyle{-\mathscr{I}_{41}} \\
 \scriptstyle{0} &  \scriptstyle{(\ell_1-k ) \mathscr{Q}_{14}} & \scriptstyle{0} & \scriptstyle{\mathscr{I}_{41}} &  \scriptstyle{(\ell_2-l ) \mathscr{Q}_{42}} & \scriptstyle{0} \\
 \scriptstyle{0} & \scriptstyle{0} &  \scriptstyle{(l-\ell_2 ) \mathscr{Q}_{42}} & \scriptstyle{\mathscr{I}_{42}} & \scriptstyle{0} &  \scriptstyle{(\ell_1-k ) \mathscr{Q}_{14}} \\
 \scriptstyle{k \mathscr{Q}_{14}} & \scriptstyle{0} & \scriptstyle{-\mathscr{I}_{41}} & \scriptstyle{0} & \scriptstyle{0} & \scriptstyle{\mathscr{I}_{42}} \\
 \scriptstyle{-l \mathscr{Q}_{42}} & \scriptstyle{-\mathscr{I}_{42}} & \scriptstyle{0} & \scriptstyle{0} & \scriptstyle{-\mathscr{I}_{41}} & \scriptstyle{0} \\
 \scriptstyle{0} & \scriptstyle{k \mathscr{Q}_{14}} & \scriptstyle{0} & \scriptstyle{-\mathscr{I}_{41}} & \scriptstyle{0} & \scriptstyle{-l \mathscr{Q}_{42}} \\
 \scriptstyle{0} & \scriptstyle{0} & \scriptstyle{-l \mathscr{Q}_{42}} & \scriptstyle{-\mathscr{I}_{42}} & \scriptstyle{-k \mathscr{Q}_{14}} & \scriptstyle{0}
\end{array}
\right)\nonumber,
\end{eqnarray}
with
\begin{eqnarray}
\check{\mathscr{Y}}^{k,l}_n\equiv
\left(\begin{array}{cccccccccccc}
  \scriptstyle{\mathscr{Y}^{k,l;1}_{n;1}} & \scriptstyle{\mathscr{Y}^{k,l;1}_{n;2}} & \scriptstyle{\mathscr{Y}^{k,l;1}_{n;3}} & \scriptstyle{\mathscr{Y}^{k,l;1}_{n;4}} & \scriptstyle{0} & \scriptstyle{0} & \scriptstyle{0} & \scriptstyle{0}  \\
  \scriptstyle{\mathscr{Y}^{k,l;2}_{n;1}} & \scriptstyle{\mathscr{Y}^{k,l;2}_{n;2}} & \scriptstyle{\mathscr{Y}^{k,l;2}_{n;3}} & \scriptstyle{\mathscr{Y}^{k,l;2}_{n;4}} & \scriptstyle{0} & \scriptstyle{0} & \scriptstyle{0} & \scriptstyle{0}  \\
  \scriptstyle{\mathscr{Y}^{k,l;3}_{n;1}} & \scriptstyle{\mathscr{Y}^{k,l;3}_{n;2}} & \scriptstyle{\mathscr{Y}^{k,l;3}_{n;3}} & \scriptstyle{\mathscr{Y}^{k,l;3}_{n;4}} & \scriptstyle{0} & \scriptstyle{0} & \scriptstyle{0} & \scriptstyle{0}  \\
  \scriptstyle{0} & \scriptstyle{0} & \scriptstyle{0} & \scriptstyle{0} & \scriptstyle{\mathscr{Y}^{k-1,l;1}_{n-1;1}} & \scriptstyle{\mathscr{Y}^{k,l-1;1}_{n-1;2}} & \scriptstyle{\mathscr{Y}^{k-1,l;1}_{n-1;3}} & \scriptstyle{\mathscr{Y}^{k,l-1;1}_{n-1;4}} \\
  \scriptstyle{0} & \scriptstyle{0} & \scriptstyle{0} & \scriptstyle{0} & \scriptstyle{\mathscr{Y}^{k-1,l;2}_{n;1}} & \scriptstyle{\mathscr{Y}^{k,l-1;2}_{n;2}} & \scriptstyle{\mathscr{Y}^{k-1,l;2}_{n;3}} & \scriptstyle{\mathscr{Y}^{k,l-1;2}_{n;4}} \\
  \scriptstyle{0} & \scriptstyle{0} & \scriptstyle{0} & \scriptstyle{0} & \scriptstyle{\mathscr{Y}^{k-1,l;3}_{n-1;1}} & \scriptstyle{\mathscr{Y}^{k,l-1;3}_{n-1;2}} & \scriptstyle{\mathscr{Y}^{k-1,l;3}_{n-1;3}} & \scriptstyle{\mathscr{Y}^{k,l-1;3}_{n-1;4}}
\end{array}\right).
\end{eqnarray}
The explicit matrix inversion gives
\begin{eqnarray}\label{eqn;SCase3}
&&\mathscr{Z}=\left(
\begin{array}{llllll}
 \frac{1}{\ell_1 \mathscr{Q} _{34}} & \frac{1}{\ell_1 \mathscr{Q} _{34} \mathscr{I} _{43}} & \scriptstyle{0} & \frac{1}{\ell_1 \mathscr{Q} _{34}} & \frac{1}{\ell_1 \mathscr{Q} _{34} \mathscr{I} _{43}} & \scriptstyle{0} \\
 \scriptstyle{0} & \scriptstyle{0} & \frac{1}{\ell_1 \mathscr{Q} _{34}} & \scriptstyle{0} & \scriptstyle{0} & \frac{1}{\ell_1 \mathscr{Q} _{34}} \\
 \frac{n}{\ell_1 \mathscr{I} _{43}} & \frac{n-\ell_1}{\ell_1 \mathscr{I} _{43}^2} & \frac{1}{\ell_1 \mathscr{Q} _{34} \mathscr{I} _{43}^2} & \frac{n-\ell_1}{\ell_1 \mathscr{I} _{43}} & \frac{n}{\ell_1 \mathscr{I}
   _{43}^2} & \frac{1}{\ell_1 \mathscr{Q} _{34} \mathscr{I} _{43}^2} \\
 \scriptstyle{0} & \scriptstyle{0} & \frac{n}{\ell_1 \mathscr{I} _{43}} & \scriptstyle{0} & \scriptstyle{0} & \frac{n-\ell_1}{\ell_1 \mathscr{I} _{43}} \\
 \scriptstyle{0} & \scriptstyle{0} & \frac{1}{\ell_1 \mathscr{Q} _{43} \mathscr{I} _{43}} & \scriptstyle{0} & \scriptstyle{-}\frac{1}{\mathscr{I} _{43}} & \frac{1}{\ell_1 \mathscr{Q} _{43} \mathscr{I} _{43}} \\
 \scriptstyle{0} & \scriptstyle{-}\frac{1}{\mathscr{I} _{43}} & \frac{1}{\ell_1 \mathscr{Q} _{34} \mathscr{I} _{43}} & \scriptstyle{0} & \scriptstyle{0} & \frac{1}{\ell_1 \mathscr{Q} _{34} \mathscr{I} _{43}}
\end{array}
\right)\check{\mathscr{Y}}^{k,l}_n\times\\
&&\qquad\qquad\qquad\qquad\times
\left(
\begin{array}{cccccc}
  \scriptstyle{(\ell_1-k ) \mathscr{Q}_{14}} & \scriptstyle{0} & \scriptstyle{\mathscr{I}_{41}} & \scriptstyle{0} & \scriptstyle{\mathscr{I}_{42}} & \scriptstyle{0} \\
  \scriptstyle{(l-\ell_2 ) \mathscr{Q}_{42}} & \scriptstyle{\mathscr{I}_{42}} & \scriptstyle{0} & \scriptstyle{0} & \scriptstyle{0} & \scriptstyle{-\mathscr{I}_{41}} \\
 \scriptstyle{0} &  \scriptstyle{(\ell_1-k ) \mathscr{Q}_{14}} & \scriptstyle{0} & \scriptstyle{\mathscr{I}_{41}} &  \scriptstyle{(\ell_2-l ) \mathscr{Q}_{42}} & \scriptstyle{0} \\
 \scriptstyle{0} & \scriptstyle{0} &  \scriptstyle{(l-\ell_2 ) \mathscr{Q}_{42}} & \scriptstyle{\mathscr{I}_{42}} & \scriptstyle{0} &  \scriptstyle{(\ell_1-k ) \mathscr{Q}_{14}} \\
 \scriptstyle{k \mathscr{Q}_{14}} & \scriptstyle{0} & \scriptstyle{-\mathscr{I}_{41}} & \scriptstyle{0} & \scriptstyle{0} & \scriptstyle{\mathscr{I}_{42}} \\
 \scriptstyle{-l \mathscr{Q}_{42}} & \scriptstyle{-\mathscr{I}_{42}} & \scriptstyle{0} & \scriptstyle{0} & \scriptstyle{-\mathscr{I}_{41}} & \scriptstyle{0} \\
 \scriptstyle{0} & \scriptstyle{k \mathscr{Q}_{14}} & \scriptstyle{0} & \scriptstyle{-\mathscr{I}_{41}} & \scriptstyle{0} & \scriptstyle{-l \mathscr{Q}_{42}} \\
 \scriptstyle{0} & \scriptstyle{0} & \scriptstyle{-l \mathscr{Q}_{42}} & \scriptstyle{-\mathscr{I}_{42}} & \scriptstyle{-k \mathscr{Q}_{14}} & \scriptstyle{0}
\end{array}
\right).\nonumber
\end{eqnarray}
It is now straightforward to do the matrix multiplication. This
solves the final case. Once again, the dependence of the entries
solely on the difference of the spectral parameters, and on the
characteristic combinations of representation labels already
observed in Case II, is a noticeable feature of the result.

\section{Reduction and Comparison}\label{sect;Reduction}

Let us now compare our formulae with the known S-matrices. Here,
one runs into potential difficulties. The formulae from the
previous sections were derived for generic bound states, and one
might wonder whether there could be obstructions for small bound
states. A first problem arises when $n$ is comparable to
$\ell_1,\ell_2$. A second problem is encountered for $n=0,n=k+l$,
since the basis of two-particle states in these two cases is
lower-dimensional. One can wonder whether our formulae
\begin{eqnarray}
\S\stateA{k,l} &=& \sum_{n=0}^{k+l}\mathscr{X}^{k,l}_{n}\stateA{n,N-n}\\
\S\stateB{k,l}_i &=& \sum_{n=0}^{k+l}\sum_{j=1}^4\mathscr{Y}^{k,l;j}_{n;i}\stateB{n,N-n}_j\label{eqn;reduction}\\
\S\stateC{k,l}_i &=&
\sum_{n=0}^{k+l}\sum_{j=1}^6\mathscr{Z}^{k,l;j}_{n;i}\stateC{n,N-n}_j,
\end{eqnarray}
with $N=k+l$ and $\mathscr{Y},\mathscr{Z}$ given by
(\ref{eqn;SCase2}) and (\ref{eqn;SCase3}), remain valid also for
these particular values.

It turns out that this is indeed the case. Let us deal with the
first problem. One can see from (\ref{eqn;SCase1}) that, when
$n>\ell_1$, precisely the unwanted S-matrix elements vanish,
basically thanks to the vanishing of the correspondent
coefficients $\mathscr{X}^{k,l}_{n}$.

Concerning the second potential problem, we notice that the issue
arises only for Case II and III states. In Case II, the
corresponding sum on the right hand side of (\ref{eqn;reduction})
contains terms like
\begin{eqnarray}
\mathscr{Y}^{k,l;4}_{0;i}\stateB{0,N}_4.
\end{eqnarray}
But, as seen from (\ref{eqn;BasisCase2}), $\stateB{0,N}_4$ is not
well-defined (actually it is not part of our bound state
representation). Hence, the S-matrix transition amplitudes toward
these states, $\mathscr{Y}^{k,l;4}_{0;i}$, should vanish
identically. We verified that this indeed turns out to be the
case, which means that these states completely decouple.

More specifically, from (\ref{eqn;SCase1}) it can be shown that
\begin{eqnarray}
\mathscr{X}^{k+1,l-1}_0 &=&\frac{\ell_2-l}{\delta u-\frac{\ell_1-\ell_2}{2}-l+1} \mathscr{X}^{k,l}_0\\
\mathscr{X}^{k-1,l+1}_0 &=&\frac{\delta u-\frac{\ell_1-\ell_2}{2}-l}{\ell_2-l-1} \mathscr{X}^{k,l}_0.
\end{eqnarray}
This means that in (\ref{eqn;SCase2}) one can pull out a factor
$\mathscr{X}^{k,l}_0$. The remaining matrix part is
straightforwardly seen to have zeroes for the states corresponding
to the amplitudes $\mathscr{Y}^{k,l;4}_{0;i}$, for all $i=1,2,3,4$
as indeed should be the case. In other words, one can
unambiguously write
\begin{eqnarray}
\S \stateB{k,l}_{i} = \sum_{n=0}^{k+l}\sum_{j=1}^{4} \mathscr{Y}^{k,l;j}_{n;i}\stateB{n,N-n}_j,
\end{eqnarray}
where $\mathscr{Y}^{k,l;j}_{n;i}$ is given by the complete
$4\times4$ matrix from (\ref{eqn;SCase2}). The same should be true
for Case III states.

One can now compare our coefficients against the known S-matrices.
Complete agreement is found with $\S^{AA},\S^{AB},\S^{BB}$ from
\cite{Beisert:2005tm,Arutyunov:2006yd,Arutyunov:2008zt}. We also
checked several coefficients of the S-matrix $\S^{1\ell}$ from
\cite{Bajnok:2008bm} which also agree with our findings.


\section*{Acknowledgements}

We thank Sergey Frolov for many valuable discussions. One of us (A.T.) wishes to thank Davide Fioravanti, George Jorjadze, Peter Orland and Jan Plefka for discussions. We are also grateful to Rafael Nepomechie for useful comments and pointing out several typos in the manuscript. The work of G.~A. was supported in part by the RFBI grant 08-01-00281-a, by the grant NSh-672.2006.1, by NWO grant 047017015 and by the INTAS contract 03-51-6346.

\appendix

\section{Yangians and Coproducts}\label{App;Yangian}

The double Yangian $DY(\mathfrak{g})$ of a (simple) Lie algebra
$\mathfrak{g}$ is a deformation of the universal enveloping
algebra $U(\mathfrak{g}[u,u^{-1}])$ of the loop algebra
$\mathfrak{g}[u,u^{-1}]$. The Yangian is generated by level $n$
generators $\mathbb{J}^{A}_{n},\ n\in\mathbb{Z}$ that satisfy the
commutation relations
\begin{eqnarray}
\ [\mathbb{J}^{A}_{m},\mathbb{J}^{B}_{n}  ]  = F^{AB}_{C}
\mathbb{J}^{C}_{m+n} + \mathcal{O}(\hbar),
\end{eqnarray}
where $F^{AB}_{C}$ are the structure constants of $\mathfrak{g}$.
The level-0 generators $\mathbb{J}_{0}^{A}$ span the Lie-algebra.
The $\su(2|2)$ Yangian has the following coproduct
\cite{Gomez:2006va,Plefka:2006ze,Beisert:2007ds}:
\begin{eqnarray}\label{eqn;CoProdYang}
\Delta (\mathbb{J}^{A}_{n})  &=& \mathbb{J}^{A}_{n}\otimes
\mathbbm{1} + \mathcal{U}^{[A]}\otimes \mathbb{J}^{A}_{n} +
\frac{\hbar }{2} \sum_{m=0}^{n-1} F_{BC}^{A}
\mathbb{J}^{B}_{n-1-m}\mathcal{U}^{[C]}\otimes \mathbb{J}^{C}_{m} +\mathcal{O}(\hbar^{2}),\nonumber\\
\Delta(\mathcal{U})&=&\mathcal{U}\otimes \mathcal{U},
\end{eqnarray}
where $\mathcal{U}$, the `braiding factor', equals $e^{i
\frac{p}{2}}$, and $\hbar = 1/g$.\smallskip

An important representation of the Yangian is the evaluation
representation. This representation consists of states
$|u\rangle$, with action $\mathbb{J}^{A}_{n}|u\rangle =
u^{n}\mathbb{J}^{A}_{0}|u\rangle$. In this representation the
coproduct structure is fixed in terms of the coproducts of
$\mathbb{J}_0,\mathbb{J}_1$. For the remainder of this paper we
will work in this representation, and identify
$\mathbb{J}_1\equiv\hat{\mathbb{J}} = u\mathbb{J}$ for the
$\su(2|2)$ Yangian.  The spectral parameter $u$ depends on
$x^{\pm}$, as one can see from formula (\ref{eqn;defu}).\smallskip

The S-matrix is a map between the following representations:
\begin{eqnarray}
\S:~~ \V_{\ell_{1}}(p_{1},e^{ip_{2}})\otimes \V_{\ell_{2}}(p_{2},1
)\longrightarrow \V_{\ell_{1}}(p_{1},1 )\otimes
\V_{\ell_{2}}(p_{2},e^{ip_{1}}),
\end{eqnarray}
where $\V_{\ell_{i}}(p_{i},e^{2i\xi})$ is the $\ell_{i}$-bound
state representation with parameters $a_{i},b_{i},c_{i},d_{i}$
with the choice of $\zeta = e^{2i\xi}$. This specific choice
removes the braiding factor $\mathcal{U}$ from appearing
explicitly in the formulas \cite{Arutyunov:2006yd}.\smallskip

The bound state S-matrices are now fixed, up to an overall phase,
by requiring invariance under the coproducts of the (Yangian)
symmetry generators
\begin{eqnarray}
\S~\Delta(\mathbb{J}^{A})&=&\Delta^{op}(\mathbb{J}^{A})~\S,\nonumber\\
\S~\Delta(\hat{\mathbb{J}}^{A})&=&\Delta^{op}(\hat{\mathbb{J}}^{A})~\S,
\end{eqnarray}
where $\Delta^{op} = P \Delta$, with $P$ the graded permutation.
\smallskip

For completeness and future reference, we give the explicit
formulas for the coproducts and for the parameters
$a_i,b_i,c_i,d_i$. First, the $\su(2|2)$ operators:
\begin{eqnarray}\label{eqn;CoprodSymm}
\Delta(\mathbb{J}_{0}^{A}) =\mathbb{J}_{1;0}^{A} +
\mathbb{J}_{2;0}^{A}.
\end{eqnarray}
The coproducts of the Yangian generators are then given by
\cite{Beisert:2007ds}
\begin{eqnarray}\label{eqn;CoProdYangDiff}
\Delta(\hat{\mathbb{L}}^{a}_{\ b}) &=& \hat{\mathbb{L}}^{\
a}_{1;b} + \hat{\mathbb{L}}^{\ a}_{2;b} + \frac{1}{2}\mathbb{L}^{\
c}_{1;b}\mathbb{L}^{\ a}_{2;c}-\frac{1}{2} \mathbb{L}^{\
a}_{1;c}\mathbb{L}^{\ c}_{2;b}-\frac{1}{2} \mathbb{G}^{\
\gamma}_{1;b}\mathbb{Q}^{\ a}_{2;\gamma}-\frac{1}{2} \mathbb{Q}^{\
a}_{1;\gamma}\mathbb{G}^{\
\gamma}_{2;b}\nonumber\\
&&+\frac{1}{4}\delta^{a}_{b}\mathbb{G}^{\
\gamma}_{1;c}\mathbb{Q}^{\ c}_{2;\gamma}+\frac{1}{4}\delta^{a}_{b}
\mathbb{Q}^{\ c}_{1;\gamma}\mathbb{G}^{\
\gamma}_{2;c}~,\nonumber\\
\Delta(\hat{\mathbb{R}}^{\alpha}_{\ \beta}) &=&
\hat{\mathbb{R}}^{\ \alpha}_{1;\beta} + \hat{\mathbb{R}}^{\
\alpha}_{2;\beta} - \frac{1}{2}\mathbb{R}^{\
\gamma}_{1;\beta}\mathbb{R}^{\ \alpha}_{2;\gamma}+\frac{1}{2}
\mathbb{R}^{\ \alpha}_{1;\gamma}\mathbb{R}^{\
\gamma}_{2;\beta}+\frac{1}{2} \mathbb{G}^{\
\alpha}_{1;c}\mathbb{Q}^{\ c}_{2;\beta}+\frac{1}{2} \mathbb{Q}^{\
c}_{1;\beta}\mathbb{G}^{\
\alpha}_{2;c}\nonumber\\
&&-\frac{1}{4}\delta^{\alpha}_{\beta}\mathbb{G}^{\
\gamma}_{1;c}\mathbb{Q}^{\
c}_{2;\gamma}-\frac{1}{4}\delta^{\alpha}_{\beta} \mathbb{Q}^{\
c}_{1;\gamma}\mathbb{G}^{\
\gamma}_{2;c}~,\\
\Delta(\hat{\mathbb{Q}}^{a}_{\ \beta}) &=& \hat{\mathbb{Q}}^{\
a}_{1;\beta} + \hat{\mathbb{Q}}^{\ a}_{2;\beta} -
\frac{1}{2}\mathbb{R}^{\ \gamma}_{1;\beta}\mathbb{Q}^{\
a}_{2;\gamma}+\frac{1}{2} \mathbb{Q}^{\ a}_{1;\gamma}\mathbb{R}^{\
\gamma}_{2;\beta} -\frac{1}{2} \mathbb{L}^{\ a}_{1;c}\mathbb{Q}^{\
c}_{2;\beta}+\frac{1}{2} \mathbb{Q}^{\
c}_{1;\beta}\mathbb{L}^{\ a}_{2;c}\nonumber\\
&&-\frac{1}{4}\mathbb{H}_{1}\mathbb{Q}^{\
a}_{2;\beta}+\frac{1}{4}\mathbb{Q}^{\ a}_{1;\beta}\mathbb{H}_{2} +
\frac{1}{2}\epsilon_{\beta\gamma}\epsilon^{ad}\mathbb{C}_{1}\mathbb{G}^{\
\gamma}_{2;d}-\frac{1}{2}\epsilon_{\beta\gamma}\epsilon^{ad}\mathbb{G}^{\
\gamma}_{1;d}\mathbb{C}_{2}~,\nonumber\\
\Delta(\hat{\mathbb{G}}^{\alpha}_{\ b}) &=& \hat{\mathbb{G}}^{\
\alpha}_{1;b} + \hat{\mathbb{G}}^{\ \alpha}_{2;b} +
\frac{1}{2}\mathbb{L}^{\ c}_{1;b}\mathbb{G}^{\ \alpha}_{2;c}-
\frac{1}{2}\mathbb{G}^{\ \alpha}_{1;c}\mathbb{L}^{\ c}_{2;b}
+\frac{1}{2} \mathbb{R}^{\ \alpha}_{1;\gamma}\mathbb{G}^{\
\gamma}_{2;b} -\frac{1}{2} \mathbb{G}^{\
\gamma}_{1;b}\mathbb{R}^{\ \alpha}_{2;\gamma}\nonumber\\
&&+\frac{1}{4}\mathbb{H}_{1}\mathbb{G}^{\ \alpha}_{2;b}-
\frac{1}{4}\mathbb{G}^{\ \alpha}_{1;b}\mathbb{H}_{2} -
\frac{1}{2}\epsilon_{bc}\epsilon^{\alpha\gamma}\mathbb{C}^{\dag}_{1}\mathbb{Q}^{\
c}_{2;\gamma}
+\frac{1}{2}\epsilon_{bc}\epsilon^{\alpha\gamma}\mathbb{Q}^{\
c}_{1;\gamma}\mathbb{C}^{\dag}_{2}~,\nonumber
\end{eqnarray}
and for the central charges
\begin{eqnarray}
\Delta(\hat{\mathbb{H}}) &=& \hat{\mathbb{H}}_{1} +\hat{\mathbb{H}}_{2}+ \mathbb{C}_{1}\mathbb{C}^{\dag}_{2}-\mathbb{C}^{\dag}_{1}\mathbb{C}_{2},\nonumber\\
\Delta(\hat{\mathbb{C}}) &=& \hat{\mathbb{C}}_{1}+\hat{\mathbb{C}}_{2}+\frac{1}{2} \mathbb{H}_{1}\mathbb{C}_{2}-\frac{1}{2} \mathbb{C}_{1}\mathbb{H}_{2},\\
\Delta(\hat{\mathbb{C}}^{\dag}) &=&
\hat{\mathbb{C}}^{\dag}_{1}+\hat{\mathbb{C}}^{\dag}_{2}+\frac{1}{2}
\mathbb{H}_{1}\mathbb{C}^{\dag}_{2}-\frac{1}{2}
\mathbb{C}^{\dag}_{1}\mathbb{H}_{2}.\nonumber
\end{eqnarray}
The product is ordered, e.g. $\mathbb{Q}_{1}\mathbb{Q}_{2}$ means
first applying $\mathbb{Q}_{2}$, then $\mathbb{Q}_{1}$ (as
differential operators). Finally, the coefficients used in
$\Delta$ are given by:
\begin{eqnarray}
\begin{array}{lll}
  a_{1} = \sqrt{\frac{g}{2\ell_{1}}}\eta_{1}, & ~ & b_{1} =
 i e^{i p_{2}}\sqrt{\frac{g}{2\ell_{1}}}~
\frac{1}{\eta_{1}}\left(\frac{x_{1}^{+}}{x_{1}^{-}}-1\right), \\
  c_{1} = -e^{-i p_{2}}\sqrt{\frac{g}{2\ell_{1}}}\frac{\eta_{1}}{ x_{1}^{+}}, & ~ &
d_{1}=i\sqrt{\frac{g}{2\ell_{1}}}\frac{x_{1}^{+}}{\eta_{1}}\left(\frac{x_{1}^{-}}{x_{1}^{+}}-1\right),\\
\eta_{1} =
e^{i\frac{p_{1}}{4}}e^{i\frac{p_{2}}{2}}\sqrt{ix^{-}_{1}-ix^{+}_{1}},&~&~\\
~ &~& ~ \\
  a_{2} = \sqrt{\frac{g}{2\ell_{2}}}\eta_{2}, & ~ & b_{2} = i\sqrt{\frac{g}{2\ell_{2}}}
\frac{1}{\eta_{2}}\left(\frac{x_{2}^{+}}{x_{2}^{-}}-1\right), \\
  c_{2} = -\sqrt{\frac{g}{2\ell_{2}}}\frac{\eta_{2}}{x_{2}^{+}}, & ~ &
d_{2}=i\sqrt{\frac{g}{2\ell_{2}}}\frac{x_{2}^{+}}{\eta_{2}}\left(\frac{x_{2}^{-}}{x_{2}^{+}}-1\right),\\
\eta_{2} = e^{i\frac{p_{2}}{4}}\sqrt{ix^{-}_{2}-ix^{+}_{2}}.&~&~
\end{array}
\end{eqnarray}
The coefficients in $\Delta^{op}$ are given by:
\begin{eqnarray}
\begin{array}{lll}
  a_{3} = \sqrt{\frac{g}{2\ell_{1}}}\eta^{op}_{1}, & ~ & b_{3} = i\sqrt{\frac{g}{2\ell_{1}}}
\frac{1}{\eta^{op}_{1}}\left(\frac{x_{1}^{+}}{x_{1}^{-}}-1\right), \\
  c_{3} = -\sqrt{\frac{g}{2\ell_{1}}}\frac{\eta^{op}_{1}}{x_{1}^{+}}, & ~ &
d_{3}=i\sqrt{\frac{g}{2\ell_{1}}}\frac{x_{1}^{+}}{\eta^{op}_{1}}\left(\frac{x_{1}^{-}}{x_{1}^{+}}-1\right),\\
\eta^{op}_{1} =
e^{i\frac{p_{1}}{4}}\sqrt{ix^{-}_{1}-ix^{+}_{1}},&~&~\nonumber\\
 ~ &~& ~ \\
  a_{4} = \sqrt{\frac{g}{2\ell_{2}}}\eta^{op}_{2}, & ~ &
b_{4} =
 i e^{i p_{1}}\sqrt{\frac{g}{2\ell_{2}}}~
\frac{1}{\eta^{op}_{2}}\left(\frac{x_{2}^{+}}{x_{2}^{-}}-1\right), \\
  c_{4} = -e^{-i p_{1}}\sqrt{\frac{g}{2\ell_{2}}}\frac{\eta^{op}_{2}}{ x_{2}^{+}}, & ~ &
d_{4}=i\sqrt{\frac{g}{2\ell_{2}}}\frac{x_{2}^{+}}{\eta^{op}_{2}}\left(\frac{x_{2}^{-}}{x_{2}^{+}}-1\right),\\
\eta^{op}_{2} =
e^{i\frac{p_{2}}{4}}e^{i\frac{p_{1}}{2}}\sqrt{ix^{-}_{2}-ix^{+}_{2}}.&~&~
\end{array}
\end{eqnarray}
The non-trivial braiding factors are all hidden in the parameters
of the four representations involved.

\section{Poles of the general Case I Amplitude}\label{sect;AppPoles}
Let us now turn to formula (\ref{eqn;SCase1}), and outline a proof
that it also does not have poles at physical values of the
momenta. In order to do this, we will make use of its rewriting in
terms of the $6j$-symbol according to formulas (\ref{6jsy}),
(\ref{co6j}). The advantage is that the $6j$-symbol itself is
analytic, and the singularities are essentially factored out in
terms of gamma functions.

First, by looking at (\ref{eqn;SCase1}), and remembering the
discussion of the case $l=0$, we see that the source of possible
physical poles is the denominator $1/\prod_{p=1}^{k+l}(\delta u
+\frac{\ell_1+\ell_2}{2}-p)$. In order to have poles at a positive
(nonzero) bound-state rank we need (see Section
\ref{sect:EasyPoles}) $\delta u = p - [(\ell_1 + \ell_2)/2]$, $p
\in ((\ell_1 + \ell_2)/2,k+l]$, $k+l
>(\ell_1 + \ell_2)/2$. This denominator is recognized as the ratio
of gamma functions $\Gamma
\left(l+\frac{\ell_1-\ell_2}{2}-n-\delta u \right)/\Gamma
\left(k+l-\frac{\ell_1+\ell_2}{2}-\delta u +1\right)$ in
(\ref{hypergeom}). From (\ref{eqn;SCase1}), it is clear that the
only thing that can happen is the rest of the formula cancelling
some of these poles with zeros. Let us first analyze just the
contribution coming from the hypergeometric function, rewritten as
in (\ref{6jsy}). After identifying the values of the parameters,
namely
\begin{eqnarray}
&&a_1 = - k, \qquad \qquad \qquad \qquad a_2 = - n,\nonumber\\
&&a_3 = 1 - \frac{\ell_1-\ell_2}{2}  + \delta u,\nonumber\\
&&a_4 = - \frac{\ell_1-\ell_2}{2} - \delta u,\nonumber\\
&&b_1 = 1 - \ell_1, \qquad \qquad \qquad b_2 = - k - l +
\ell_2,\nonumber\\
&&b_3 = 1+l-n,
\end{eqnarray}
one can see that four of the gamma functions in the numerator and
four in the denominator of (\ref{6jsy}) bear dependence on $\delta u$.
At the values of $\delta u$ corresponding to physical
poles, two of these gamma functions in the denominator become
$\Gamma \left(-k-l-p\right)$ and $\Gamma \left(-1-k-l+\ell_1 +
\ell_2 -p\right)$. The first one has always negative argument,
since $p\leq k+l$. The second one as well, since $-1-k-l+\ell_1 +
\ell_2 - p < -1-k-l+[(\ell_1 + \ell_2)/2]<-1-2k-2l<0$, where we
have used the physical conditions on $p$ and $k+l$ reported above.
These two poles under the square root combine to give a simple
pole in the denominator, namely a zero, that cancels the physical
pole coming from the ratio of gamma functions discussed above.
However, this does not terminate our analysis, since we still have
to make sure no other physical poles are generated from all the
other gamma functions still present in the formula. Let us spell
those singularities out.

Two of the four gamma functions depending on $\delta u$ in the
numerator are always regular at the physical value of momenta.
They reduce in fact to $\Gamma \left(1+p\right)$ and $\Gamma
\left(\ell_1 + \ell_2 -p\right)$. The first one has manifestly
positive nonzero argument, and so does the second one, since
$\ell_1 + \ell_2 -p>\ell_1 + \ell_2 - k - l \geq 2$ using
conditions on $p$ and $\ell_1 \geq k+1$, $\ell_2 \geq l+1$ from
(\ref{transit}). Poles could only arise from the remaining two
gamma functions depending on $\delta u$ in the numerator, which at
the physical values of momenta become $\Gamma \left(1-\ell_2 -
p\right)$ and $\Gamma \left(\ell_1 -p\right)$. These two poles are
however cancelled by a zero coming from $\Gamma
\left(\frac{\ell_1-\ell_2}{2}-\delta u \right)$ in the denominator
in (\ref{hypergeom}). The latter reduces to $\Gamma \left(\ell_1
-p\right)$ at physical momenta. Factoring it into two identical
square roots, we see that one of them exactly cancels one
contribution from the hypergeometric function, and the other one
cancels the singularities of the other. In fact, when $\Gamma
\left(1-\ell_2 - p\right)$ has poles, namely for $1-\ell_2 + p\leq
0$, this implies $\ell_2 - 1 \geq p > (\ell_1 + \ell_2)/2$, {\it
i.e.} $(\ell_1 - \ell_2)/2 <-1$. But this means that $\ell_1
-p<(\ell_1 - \ell_2)/2 <-1$, and the two poles cancel. In order to
conclude the argument, we still need to analyze potential poles
coming from the gamma function $\Gamma
\left(l+\frac{\ell_1-\ell_2}{2}-n-\delta u \right)$ in the
numerator of (\ref{hypergeom}). But when this gamma has a pole,
namely for $l+\ell_1 - n - p<0$, one of the four gammas depending
on $u_1 - u_2$ in the denominator of formula (\ref{6jsy}) (and
different from the two already considered), becomes $\Gamma
\left(l+\ell_1 - n - p \right)$, which has also a pole. So the
latter reduces the upper pole to a square root of it. However, the
original formula does not have any branch cut, therefore it is not
possible to have one square root singularity left uncancelled.
Since one can check that all the other parts of the formula,
excluding the $6j$-symbol, are neutralized, one concludes that the
$6j$-symbol must have a zero in this case. We checked that this is
the case for few examples, where one can see that the
triangularity condition for the $6j$-symbol is violated.

\bibliographystyle{JHEP}
\bibliography{LitRmat}

\end{document}